\documentclass[aps,pra,reprint,groupedaddress,twocolumn,10pt]{revtex4-1}
\usepackage{gnuplottex}
\usepackage{mathtools}
\usepackage{lipsum}
\usepackage[usenames,dvipsnames]{xcolor}
\usepackage{float}

\newcommand{\etal}{{\it{et al.}}}

\definecolor{dgreen}{rgb}{0,.5,0}
\definecolor{dblue}{rgb}{0,0,0.75}
\definecolor{dred}{rgb}{0.5,0,.5}

\DeclarePairedDelimiter\bra{\langle}{\rvert}
\DeclarePairedDelimiter\ket{\lvert}{\rangle}
\DeclarePairedDelimiterX\braket[2]{\langle}{\rangle}{#1 \delimsize\vert #2}
\def\ddroit{{\rm d}}

\begin{document}

\preprint{}

\title{
Exact ensemble density functional theory for excited states in a model system: investigating the weight dependence of the correlation energy
}


\author{
Killian Deur, Laurent Mazouin, and Emmanuel Fromager$^{{\ast}}$\\
\footnote{$^\ast$Corresponding author. Email: fromagere@unistra.fr}
{\em{
Laboratoire de Chimie Quantique,
Institut de Chimie, CNRS/Universit\'{e} de Strasbourg,\\
4 rue Blaise Pascal, 67000 Strasbourg, France\\
}}
}


\date{\today}

\begin{abstract}


Ensemble density functional theory (eDFT) is an exact time-independent
alternative to time-dependent DFT (TD-DFT) for the calculation of excitation
energies. Despite its formal simplicity and advantages in contrast to
TD-DFT (multiple excitations, for example, can be easily taken into account in an
ensemble), eDFT is not standard which is essentially due to the lack of
reliable approximate exchange-correlation ($xc$) functionals for ensembles.
Following Smith {\it et al.} [Phys. Rev. B {\bf 93}, 245131 (2016)], we
propose in this work to construct an exact eDFT for the nontrivial
asymmetric Hubbard dimer, thus providing more insight into the 
weight dependence of the ensemble $xc$ energy in various correlation
regimes. For that purpose, an exact analytical expression for the
weight-dependent ensemble exchange energy has been derived. The
complementary exact ensemble correlation energy has been computed by
means of Legendre--Fenchel transforms. Interesting features like
discontinuities in the ensemble $xc$ potential in the strongly
correlated limit have been rationalized by means of a generalized
adiabatic connection formalism. Finally, functional-driven errors
induced by ground-state density-functional approximations have been
studied. In the strictly symmetric case or in the weakly  
correlated regime, combining ensemble exact exchange with ground-state
correlation functionals gives relatively accurate ensemble energies.
However,  
when approaching the equiensemble in the strongly correlated
regime, this approximation leads to highly curved ensemble energies with
negative slope which is unphysical. Using both  
ground-state exchange and correlation functionals gives much better
results in that case. In fact, exact ensemble energies are almost recovered in
some density domains. The analysis of density-driven errors is left for
future work. 


\end{abstract}



\maketitle


\section{Introduction}

Despite its success, time-dependent density functional theory
(TD-DFT)~\cite{PRL84_RGth}
within the adiabatic local or semi-local approximation
still suffers from various deficiencies like the underestimation of
charge transfer excitation energies or the absence of multiple electron
excitations in the spectrum~\cite{Casida_tddft_review_2012}. In order to describe excited states in the
framework of DFT, it is in principle not necessary to work within the
time-dependent regime. Various time-independent DFT approaches have been
investigated over the years, mostly at the formal
level~\cite{PRA99_Goerling_DFT_beyond_HK_th,PRL99_Nagy_DFT_individual_ES,PRL04_Burke_lack_of_HK_th_ES,PRA09_Ayers_Tind-DFT_ES,JCP09_Ziegler_relation_TD-DFT_VDFT,PRA12_Nagy_TinD-DFT_ES,JCTC13_Ziegler_SCF-CV-DFT}.
In this paper, we will focus on ensemble DFT (eDFT) for excited
states~\cite{JPC79_Theophilou_equi-ensembles,PRA_GOK_EKSDFT}. The latter relies on the extension
of the variational principle to an ensemble of ground and excited states
which is characterized by a set of ensemble
weights~\cite{PRA_GOK_RRprinc}. Note that Boltzmann
weights can be used~\cite{PRA13_Pernal_srEDFT} but it is not compulsory.
In fact, any set of ordered weights can
be considered~\cite{PRA_GOK_RRprinc}. Since the ensemble
energy ({\it i.e.} the weighted sum of ground- and excited-state
energies) is a functional of the ensemble density, which is the
weighted sum of ground- and excited-state densities, a mapping between
the physical interacting and Kohn--Sham (KS) non-interacting ensembles can
be established. Consequently, a weight-dependent ensemble
exchange-correlation ($xc$) functional must be introduced in order to obtain
the exact ensemble energy and, consequently, exact excitation
energies. Despite its formal simplicity (exact optical and KS gaps are easily
related in this context~\cite{PRA_GOK_EKSDFT}) and advantages in contrast to
TD-DFT (it is straightforward to describe multiple excitations with an
ensemble), eDFT is not standard essentially because, so far, not much
effort has been
put in the development of
approximate $xc$ functionals for ensembles. In particular, designing density-functional
approximations that remove the so-called "ghost interaction"
error~\cite{PRL02_Gross_spurious_int_EDFT},
which is induced by the ensemble Hartree energy, is
still challenging~\cite{JCP14_Pernal_ghost_interaction_ensemble}. Employing an ensemble exact exchange energy is of course
possible but then optimized effective potentials should in principle be
used, which is computationally demanding. Recently, accurate eDFT
calculations have been performed for the helium
atom~\cite{PRA14_Burke_exact_GOK-DFT}, the hydrogen
molecule~\cite{AIP15_Borgoo_eDFT_H2}, and for two electrons in boxes or in a three-dimensional harmonic
well (Hooke's atom)~\cite{JCP14_Burke_GOK-DFT_two-electron-systems},
thus providing more insight into the ensemble $xc$ energy and
potential. The key feature of the $xc$ density functional in eDFT is
that it varies with the ensemble weight, even if the electron density is
fixed. This weight dependence plays a crucial role in the calculation of
the excitation energies~\cite{PRA_GOK_EKSDFT}. Developing  
weight-dependent functionals is a complicated task that has not drawn
much attention so far. This explains why
eDFT is not a standard approach. There is clearly a need for models that can be solved
exactly in eDFT and, consequently, that can provide more insight into
the weight dependence of ensemble $xc$ energies.\\  

It was shown very recently~\cite{carrascal2015hubbard,PRB16_Burke_thermal_DFT_dimer} that the nontrivial
asymmetric Hubbard dimer can be used for understanding the limitations
of standard approximate DFT in the strongly correlated regime and also for developing $xc$
functionals in thermal DFT~\cite{PRB16_Burke_thermal_DFT_dimer}. In the same spirit, we propose in this work
to construct an exact eDFT for this model system. The paper is organized
as follows. After a brief introduction to eDFT (Sec.~\ref{subsec:eDFT}),
a generalization of the adiabatic connection formalism to ensembles will
be presented in Sec.~\ref{subsec:gace}. The formulation of eDFT for the
Hubbard dimer is discussed in Sec.~\ref{sec:edft_for_Hubb_dimer} and
exact results are given and analyzed in Sec.~\ref{sec:results}.
Ground-state density-functional approximations are finally proposed and
tested in Sec.~\ref{sec:GS_approx}. Conclusions are given in
Sec.~\ref{sec:conclu}.    

\section{Theory}

\subsection{Ensemble density functional theory for excited
states}\label{subsec:eDFT}

According to the Gross--Oliveira--Kohn (GOK) variational principle~\cite{PRA_GOK_RRprinc}, that
generalizes the seminal work of
Theophilou~\cite{JPC79_Theophilou_equi-ensembles} on equiensembles, the
following inequality 
\begin{eqnarray}\label{eq:GOK_VP_Tr}
E^{\mathbf{w}}\leq 
\sum^{M-1}_{k=0}w_k\langle \overline{\Psi}_k\vert 
\hat{H}\vert\overline{\Psi}_k\rangle,
\end{eqnarray}
is fulfilled for any ensemble 
characterized by an arbitrary set ({\it i.e.} not necessarily a
Boltzmann one) of
weights $\mathbf{w}\equiv(w_0,w_1,\ldots,w_{M-1})$ with 
$w_0\geq w_1\geq\ldots\geq w_{M-1}>0$ and a set of $M$ 
orthonormal trial
 $N$-electron (with $N$ fixed) wavefunctions
$\{\overline{\Psi}_k\}_{0\leq k\leq M-1}$.  
The lower bound in Eq.~(\ref{eq:GOK_VP_Tr}) is the exact ensemble
energy, {\it i.e.} the weighted sum of ground- and excited-state
energies, 
\begin{eqnarray}
E^{\mathbf{w}}=\sum_{k=0}^{M-1}w_k\langle\Psi_k\vert\hat{H}\vert\Psi_k\rangle
=\sum_{k=0}^{M-1}w_kE_k,
\end{eqnarray} 
where $\Psi_k$ is the exact $k$th eigenfunction of the Hamiltonian
operator $\hat{H}$ with energy $E_k$ and
$E_0\leq E_1\leq \ldots\leq E_{M-1}$. A consequence of the GOK principle
is that the ensemble energy is a functional of the ensemble
density~\cite{PRA_GOK_EKSDFT},
{\it i.e.} the weighted sum of ground- and excited-state densities,   
\begin{eqnarray}
n^{\mathbf{w}}({\bf r})=\sum_{k=0}^{M-1}w_k n_{{\Psi}_k}({\bf r}).
\end{eqnarray}
Note that, in the standard formulation of eDFT~\cite{PRA_GOK_EKSDFT}, the additional condition
$\sum_{k=0}^{M-1}w_k=1$ is used so that the ensemble density integrates
to the number $N$ of electrons. In the rest of this work, we will focus
on non-degenerate two-state ensembles. In the latter case, a single
weight parameter $w=w_1$ in the range $0\leq w\leq 1/2$ can be used,
since $w_0=1-w$ and $w_0\geq w_1$, so that Eq.~(\ref{eq:GOK_VP_Tr})
becomes
\begin{eqnarray}\label{eq:GOK_VP_Tr2}
E^{{w}}\leq 
{\rm Tr}\left[\hat{\gamma}^w\hat{H}\right].
\end{eqnarray}
For convenience, the trial density matrix operator
\begin{eqnarray}
\hat{\gamma}^w=
(1-w)
\vert \overline{\Psi}_0\rangle\langle
\overline{\Psi}_0\vert
+w 
\vert \overline{\Psi}_1\rangle\langle
\overline{\Psi}_1\vert,
\end{eqnarray}
where $\overline{\Psi}_0$ and $\overline{\Psi}_1$ are orthonormal, has been introduced. ${\rm Tr}$ denotes the trace and the ensemble
energy equals
\begin{eqnarray} E^{w} = (1-w)E_{0} + w E_{1}.
\label{eq:ensembleenergy_2states}
\end{eqnarray}
For any electronic system, the Hamiltonian can be decomposed as
$\hat{H}=\hat{T}+\hat{W}_{ee}+ \int d{\bf r}\,v_{ne}({\bf r})\hat{n}({\bf
r})$ where $\hat{T}$ is the kinetic energy operator, $\hat{W}_{ee}$
denotes the two-electron repulsion operator, $v_{ne}({\bf r})$ is the
nuclear potential and $\hat{n}({\bf r})$ is the density operator. Like in conventional (ground-state) DFT, the exact ensemble energy can
be expressed variationally as follows~\cite{PRA_GOK_EKSDFT},
\begin{eqnarray}\label{eq:min_n_Ew} E^{w} = \min_{n} \left\{ F^{w}[n] + \int d{\bf r}\,
v_{ne}({\bf r}) n({\bf r})\right\},\end{eqnarray}
where
\begin{eqnarray}\label{eq:general_LL_ensfun}
F^{{w}}[n]&=&\underset{\hat{\gamma}^{{w}}\rightarrow n}{\rm
min}\left\{{\rm Tr}\left[\hat{\gamma}^{{w}}(\hat{T}+\hat{W}_{
ee})\right]\right\}
\nonumber\\
&=& 
{\rm Tr}\left[\hat{\Gamma}^{{w}}[n](\hat{T}+\hat{W}_{
ee})\right]
\end{eqnarray}
is the analog of the Levy--Lieb (LL) functional for ensembles. The
minimization in Eq.~(\ref{eq:general_LL_ensfun}) is performed over all
ensemble density matrix operators with density $n$,
\begin{eqnarray}
{\rm
Tr}\left[\hat{\gamma}^{{w}}\hat{n}(\mathbf{r})\right]=n_{\hat{\gamma}^{{w}}}(\mathbf{r})=n(\mathbf{r}).
\end{eqnarray}
Note that, according to the GOK variational principle, the following
inequality is fulfilled for any local potential $v({\bf r})$, 
\begin{eqnarray}
\label{eq:GOK_variational_LF2}
E^{{{w}}}[v] \leq F^{{w}}[n]+
\int d{\bf{r}}\,v({\bf{r}})n({\bf{r}})
,  
\end{eqnarray}
where $E^{{{w}}}[v]$ is the ensemble energy of $\hat{T}+\hat{W}_{ee}+ \int
d{\bf r}\,v({\bf r})\hat{n}({\bf r})$, so that the ensemble LL functional can be rewritten as a Legendre--Fenchel
transform~\cite{Eschrig,Kutzelnigg:2006,vanLeeuwen:2003,IJQC83_Lieb_LF_transf,MP14_Manu_GACE,AIP15_Borgoo_eDFT_H2},
\begin{eqnarray}
\label{eq:legendre_fenchel_ensembleLL}
F^{w}[n] = \sup\limits_{v} \left\{ E^{w}[v] 
-\int d{\bf{r}}\,v({\bf{r}})n({\bf{r}})
\right\}.
\end{eqnarray}
Note also that, in Eq.~(\ref{eq:min_n_Ew}), the minimizing density is the exact
physical ensemble density
\begin{eqnarray}\label{eq:phys_ens_dens}
n^w({\mathbf r})=(1-w)n_{\Psi_0}({\mathbf
r})+wn_{\Psi_1}({\mathbf r}).
\end{eqnarray}\\ 
Like in standard ground-state DFT, the KS decomposition,
\begin{eqnarray}\label{eq:ens_KS_decomp} 
F^{w}[n] = T^{w}_{s}[n] + E_{Hxc}^{w}[n],\end{eqnarray}
is usually considered, where 
\begin{eqnarray}\label{eq:nonintek}
T^{w}_{s}[n]&=&\underset{\hat{\gamma}^{{w}}\rightarrow n}{\rm
min}\left\{{\rm Tr}\left[\hat{\gamma}^{{w}}\hat{T}
\right]\right\}
\nonumber\\
&=&{\rm Tr}\left[\hat{\Gamma}_s^{{w}}[n]\hat{T} \right]
\end{eqnarray}
is the non-interacting ensemble kinetic energy and 
$E_{Hxc}^{w}[n]$ is the ($w$-dependent) ensemble 
Hartree-exchange-correlation functional. Applying the GOK
principle to non-interacting systems leads to the following
Legendre--Fenchel transform,
\begin{eqnarray}
\label{eq:legendre_fenchel_ensembleTs}
T_s^{w}[n] = \sup\limits_{v} \left\{ \mathcal{E}^{KS,w}[v] 
-\int d{\bf{r}}\,v({\bf{r}})n({\bf{r}})
\right\},
\end{eqnarray}
where $\mathcal{E}^{KS,w}[v]$ is the ensemble energy of $\hat{T}+\int
d{\bf r}\,v({\bf r})\hat{n}({\bf r})$.
Combining
Eq.~(\ref{eq:min_n_Ew})
with Eq.~(\ref{eq:ens_KS_decomp}) leads to the following KS expression for
the exact ensemble energy,
\begin{eqnarray}\label{eq:min_gamma_Ew}
 E^{w} &=& \min_{{\hat{\gamma}}^{{w}}}
\Big\{ 
{\rm Tr}\left[\hat{\gamma}^{{w}}\hat{T}
\right]+ E_{Hxc}^{w}[n_{\hat{\gamma}^{{w}}}]
\nonumber\\
&&+\int d{\bf r}\,
v_{ne}({\bf r}) n_{\hat{\gamma}^{{w}}}({\bf r})\Big\}
.
\end{eqnarray}
The minimizing non-interacting ensemble density matrix in
Eq.~(\ref{eq:min_gamma_Ew}), 
\begin{eqnarray}
\hspace{-0.4cm}
\hat{\Gamma}_s^w=(1-w)
\vert {\Phi}^{KS,w}_0\rangle\langle
{\Phi}^{KS,w}_0\vert
+w 
\vert {\Phi}^{KS,w}_1\rangle\langle
{\Phi}^{KS,w}_1\vert,
\end{eqnarray}
reproduces the exact physical ensemble density,
\begin{eqnarray}
n_{\hat{\Gamma}_s^w}({\bf r})=n^w({\bf r}).
\end{eqnarray}
It is obtained by solving the self-consistent
equations~\cite{PRA_GOK_EKSDFT}
\begin{eqnarray} 
\label{eq:eKS-dft_sc_eqs}
&&\left[\hat{T} + 
\int d{\bf r} \left(v_{ne}({\bf r})+\frac{\delta E^{w}_{Hxc}[n_{\hat{\Gamma}_s^w}]}{\delta
n({\bf r})} 
\right)\hat{n}({\bf r})\right]\ket{\Phi^{KS,w}_{i}} =
\nonumber \\ 
&&\mathcal{E}^{KS,w}_{i}\ket{\Phi_i^{KS,w}}, ~ i = 0,1.\end{eqnarray}\\

As readily seen in Eq.~(\ref{eq:ensembleenergy_2states}), the exact
(neutral) excitation energy is simply the first derivative of the
ensemble energy with respect to the ensemble weight $w$,
\begin{eqnarray} 
\dfrac{dE^w}{dw}= E_{1} - E_{0}=\omega,\hspace{0.3cm} 0\leq w\leq 1/2. 
\label{eq:excitationenergy_deriv} \end{eqnarray}
Using Eq.~(\ref{eq:min_gamma_Ew}) and the Hellmann--Feynman theorem leads to
\begin{eqnarray} 
\omega&=& 
{\rm Tr}\left[\partial_w\hat{\Gamma}_s^{{w}}\hat{T}
\right]+ 
\int d{\bf r} \left(v_{ne}({\bf r})+\frac{\delta E^{w}_{Hxc}[n_{\hat{\Gamma}_s^w}]}{\delta
n({\bf r})}\right)n_{\partial_w\hat{\Gamma}_s^{{w}}}({\bf r}) 
\nonumber\\
&&+
\left.\dfrac{\partial E_{Hxc}^{\xi}[n_{\hat{\Gamma}_s^{{w}}}]}{\partial
\xi}\right|_{\xi=w}
,
\label{eq:excitationenergy_deriv2} \end{eqnarray}
where $\partial_w \hat{\Gamma}_s^w=
\vert {\Phi}^{KS,w}_1\rangle\langle
{\Phi}^{KS,w}_1\vert - 
\vert {\Phi}^{KS,w}_0\rangle\langle
{\Phi}^{KS,w}_0\vert
.
$
By using Eq.~(\ref{eq:eKS-dft_sc_eqs}),
we finally obtain
\begin{eqnarray} 
\omega&=& 
\mathcal{E}^{KS,w}_{1} - \mathcal{E}^{KS,w}_{0} + 
\left.\dfrac{\partial E_{Hxc}^{\xi}[n_{\hat{\Gamma}_s^{{w}}}]}{\partial
\xi}\right|_{\xi=w}.
\label{eq:excitationenergy_deriv3} \end{eqnarray}
If the ground and first-excited states differ by a single electron
excitation then the KS excitation energy (first term on the right-hand
side of Eq.~(\ref{eq:excitationenergy_deriv3})) becomes the
weight-dependent KS
HOMO-LUMO gap $\varepsilon^w_L-\varepsilon^w_H$. If, in addition, 
we use the decomposition 
\begin{eqnarray}\label{eq:ensHxc_decomp}
E^{w}_{Hxc}[n]=E_H[n]+E^{w}_{xc}[n],
\end{eqnarray}
where $E_H[n]$ is the conventional (weight-independent) ground-state
Hartree functional,
\begin{eqnarray}\label{eq:EH_def} E_{H}[n] = \frac{1}{2} \iint d{\bf r} d{\bf r'}
~ \frac{n({\bf r})n({\bf r'})}{\mid{{\bf r}-{\bf r'}}\mid}, \end{eqnarray}
we then recover the KS-eDFT expression for the excitation
energy~\cite{PRA_GOK_EKSDFT},
\begin{eqnarray} 
\label{eq:xcDD}\omega = \varepsilon^w_L-\varepsilon^w_H+ \Delta^{w}_{xc}, \end{eqnarray}
where $\Delta^{w}_{xc}=\left.\partial E^{\xi}_{xc}[n^w]/\partial
\xi\right|_{\xi=w}$. Interestingly, in the $w\rightarrow 0$ limit, the
excitation energy can be expressed exactly in terms of the usual
ground-state KS HOMO-LUMO gap $\varepsilon_L-\varepsilon_H$ as
\begin{eqnarray} \omega = \varepsilon_L-\varepsilon_H+\Delta^{0}_{xc}.
 \end{eqnarray}
As shown analytically by Levy~\cite{PRA_Levy_XE-N-N-1} and illustrated numerically
by Yang {\it et al.}~\cite{PRA14_Burke_exact_GOK-DFT}, $\Delta_{
xc}^{0}$ corresponds to the jump in the $xc$ potential
when moving from $w=0$ ($N$-electron ground state) to $w\rightarrow 0$
(ensemble of $N$-electron ground and
excited states). It is therefore a derivative discontinuity (DD)
contribution to the optical gap that should
not be confused with the conventional ground-state
DD~\cite{JPCLett12_Baer_curvature_frontier_orb_energies_dft,PRL13_Kronik_grand_can_ens,JCP14_Kronik_grand_can_ens,GouTou-PRA-14},
\begin{eqnarray}\label{eq:DD_fund_gap}
\Delta_{xc}=\omega_g-\left(\varepsilon_L-\varepsilon_H\right),
\end{eqnarray}
where the fundamental gap is expressed in
terms of $N-1$, $N$ and $N+1$ {\it ground-state} energies as follows,
\begin{eqnarray}\label{eq:fund_gap}
\omega_g=E_0(N-1)+E_0(N+1)-2E_0(N).
\end{eqnarray}
For simplicity, we will also refer to the weight-dependent
quantity $\Delta_{xc}^{w}$ (see Eq.~(\ref{eq:xcDD})) as DD.\\

Returning to the decomposition in Eq.~(\ref{eq:ensHxc_decomp}), the
$xc$ contribution is usually split as follows,
\begin{eqnarray}
E^{w}_{xc}[n]=E^{w}_{x}[n]+E^{w}_{c}[n],
\end{eqnarray}
where 
\begin{eqnarray}\label{eq:ens_EXX_def}
E^{w}_{x}[n]={\rm
Tr}\left[\hat{\Gamma}_s^{{w}}[n]\hat{W}_{ee}\right]-E_H[n]
\end{eqnarray}
is the exact ensemble exchange energy functional
and $\hat{\Gamma}_s^{{w}}[n]$ is the non-interacting ensemble density matrix
operator with density $n$ (see Eq.~(\ref{eq:nonintek})). Consequently,
according to Eqs.~(\ref{eq:general_LL_ensfun}), (\ref{eq:ens_KS_decomp})
and (\ref{eq:nonintek}), the ensemble correlation energy equals
\begin{eqnarray}
E^{w}_{c}[n]&=&
{\rm Tr}\left[\hat{\Gamma}^{{w}}[n](\hat{T}+\hat{W}_{
ee})\right]
\nonumber\\
&&-
{\rm Tr}\left[\hat{\Gamma}_s^{{w}}[n](\hat{T}+\hat{W}_{
ee})\right]<0.
\end{eqnarray}
 
\subsection{Generalized adiabatic connection for
ensembles}\label{subsec:gace}

In order to construct the ensemble $xc$ functional $E^{w}_{xc}[n]$ from the ground-state
one ($w=0$), Franck and Fromager~\cite{MP14_Manu_GACE} have derived a generalized
adiabatic connection for ensembles (GACE) where an integration over both the interaction strength parameter
$\lambda$ ($0\leq \lambda \leq 1$) and an ensemble weight $\xi$ in the
range $0\leq \xi\leq w$ is performed. The major difference between conventional
ACs~\cite{LANGRETH:1975p1425,GUNNARSSON:1976p1781,LANGRETH:1977p1780,SAVIN:2003p635,Nagy_ensAC}
and the GACE is that, along a GACE path, the ensemble density is held
constant and equal to $n$ when both $\lambda$ {\it and} $\xi$ vary.
Consequently, the integration over $\lambda$ can be performed in the
ground state while the deviation of the ensemble $xc$ energy from the
ground-state one is obtained when varying $\xi$ only. Formally, the GACE can
be summarized as follows. Let us consider the Schr\"{o}dinger,
\begin{eqnarray} 
&&\left(\hat{T} + \hat{W}_{ee} + 
\int d{\bf r}\,v^\xi[n]({\bf r})\hat{n}({\bf r})\right)
\ket{\Psi_{i}^{\xi}[n]}
\nonumber\\
&&= {E}^{\xi}_{i}[n]\ket{\Psi^{\xi}_{i}[n]}\end{eqnarray}
and KS   
\begin{eqnarray} 
&&\left(\hat{T} + 
\int d{\bf r}\,v^{KS,\xi}[n]({\bf r})\hat{n}({\bf r})\right)
\ket{\Phi_{i}^{KS,\xi}[n]}
\nonumber\\
&&= \mathcal{E}^{KS,\xi}_{i}[n]\ket{\Phi^{KS,\xi}_{i}[n]}\end{eqnarray}
equations where $i=0,1$. The potentials $v^\xi[n]({\bf r})$ and
$v^{KS,\xi}[n]({\bf r})$
are adjusted so that the GACE density constraint is fulfilled,
\begin{eqnarray}\label{eq:GACE_dens_constr}
n_{\hat{\Gamma}^\xi[n]}({\mathbf
r})=n_{\hat{\Gamma}^\xi_s[n]}({\mathbf r})=n({\mathbf r}),\hspace{0.2cm} 0\leq \xi\leq w,
\end{eqnarray}
where
\begin{eqnarray}
\hat{\Gamma}^\xi[n]=
(1-\xi)
\vert {\Psi}_0^{\xi}[n]\rangle\langle
{\Psi}_0^{\xi}[n]\vert
+\xi 
\vert {\Psi}_1^{\xi}[n]\rangle\langle
{\Psi}_1^{\xi}[n]\vert
\end{eqnarray}
and
\begin{eqnarray}
\hat{\Gamma}_s^\xi[n]&=&(1-\xi)
\vert {\Phi}^{KS,\xi}_0[n]\rangle\langle
{\Phi}^{KS,\xi}_0[n]\vert
\nonumber
\\
&&+\xi
\vert {\Phi}^{KS,\xi}_1[n]\rangle\langle
{\Phi}^{KS,\xi}_1[n]\vert.
\end{eqnarray}
According to Eqs.~(\ref{eq:ens_KS_decomp}) and (\ref{eq:ensHxc_decomp}),
the ensemble $xc$ energy can be expressed as
\begin{eqnarray}
E^{w}_{xc}[n]&=&E_{xc}[n]+
\int_0^w d\xi \dfrac{\partial
E^{\xi}_{xc}[n]}{\partial \xi}
\nonumber\\
&=& E_{xc}[n]+
\int_0^w d\xi \left(
\dfrac{\partial
F^{\xi}[n]}{\partial \xi}
-
\dfrac{\partial
T_s^{\xi}[n]}{\partial \xi}
\right),
\end{eqnarray}
where $E_{xc}[n]$ is the ground-state $xc$ functional.
Since $v^\xi[n]$ and $v^{KS,\xi}[n]$ are the maximizing (and
therefore stationary) potentials in the
Legendre--Fenchel transforms of Eqs.~(\ref{eq:legendre_fenchel_ensembleLL}) and
(\ref{eq:legendre_fenchel_ensembleTs}) when $w=\xi$, respectively, 
we finally obtain
\begin{eqnarray}\label{eq:int_GACE_integrand}
E^{w}_{xc}[n]&=&E_{xc}[n]+
\int_0^w d\xi\; 
\Delta^\xi_{xc}[n],
\end{eqnarray}
where the GACE integrand is simply equal to the difference in excitation
energy between the
interacting and non-interacting electronic systems whose
ensemble density with weight $\xi$ is equal to $n$: 
\begin{eqnarray}\label{eq:general_gace_integrand}
\hspace{-0.4cm}\Delta^\xi_{xc}[n]=\left(E^{\xi}_{1}[n]-E^{\xi}_{0}[n]\right)-\left(\mathcal{E}^{KS,\xi}_{1}[n]-\mathcal{E}^{KS,\xi}_{0}[n]\right). 
\end{eqnarray}
Note that, when the density $n$ equals the physical ensemble density
$n^w$ (see Eq.~(\ref{eq:phys_ens_dens})) and $\xi=w$, the GACE integrand
equals the
$xc$ DD $\Delta^{w}_{xc}$ introduced in
Eq.~(\ref{eq:xcDD}).\\

An open and critical question is whether the GACE can actually be
constructed for all weights $\xi$ in $0\leq\xi\leq w$ and densities of
interest. In other words, does the GACE density constraint lead to
interacting and/or non-interacting $v$-representability problems ? So
far, the GACE has been constructed only for the simple hydrogen molecule
in a minimal basis and near the dissociation limit~\cite{MP14_Manu_GACE}, which basically
corresponds to the strongly correlated symmetric Hubbard dimer. In the following, we
extend this work to the nontrivial asymmetric Hubbard dimer. An
important feature of such a model is that, in contrast to the symmetric
case, the density (which is simply a collection of two site occupations) can vary, thus allowing for the construction of
density
functionals~\cite{carrascal2015hubbard,PRB16_Burke_thermal_DFT_dimer}.

\section{Asymmetric Hubbard dimer}\label{sec:edft_for_Hubb_dimer}


In the spirit of recent works by Carrascal {\it et al.}~\cite{carrascal2015hubbard} as well as
Senjean~\etal~\cite{MP16_Senjean_2L-ILDA}, we propose 
to apply eDFT to the asymmetric two-electron Hubbard dimer. The
corresponding model  
Hamiltonian is decomposed as follows,
\begin{eqnarray}\label{eq:asym_Hubb_ham}
\hat{\mathcal{H}}=\hat{\mathcal{T}}+\hat{U}
+v_0\hat{n}_{0}+v_1\hat{n}_{1},
\end{eqnarray} 
where the two sites are labelled as 0 and 1, and $\hat{\mathcal{T}}=
-t\sum_{\sigma=\uparrow,\downarrow}\Big(
\hat{a}^\dagger_{0\sigma}\hat{a}_{1\sigma}+
\hat{a}^\dagger_{1\sigma}\hat{a}_{0\sigma}
\Big)
$ is the hopping operator ($t>0$) which plays the role of the kinetic
energy operator. The two-electron repulsion becomes an on-site
repulsion,
\begin{eqnarray}
\hat{U}=U\sum^1_{i=0}\hat{n}_{i\uparrow}\hat{n}_{i\downarrow},
\end{eqnarray}   
where $\hat{n}_{i\sigma}=\hat{a}^\dagger_{i\sigma}\hat{a}_{i\sigma}$ is the spin-occupation operator. The last
two contributions on the right-hand side of Eq.~(\ref{eq:asym_Hubb_ham})
play the role of the local nuclear potential. In this context, the
density operator is
$\hat{n}_{i}=\sum_{\sigma=\uparrow,\downarrow}\hat{n}_{i\sigma}$.  
For convenience, we will
assume that 
\begin{eqnarray}\label{eq:pot_cond}
v_0+v_1=0.
\end{eqnarray}
Note that the latter condition is fulfilled by any potential once it has
been shifted by $-(v_0+v_1)/2$. Therefore, the final expression for the
Hamiltonian is  
\begin{eqnarray}\label{eq:final_ham_Hubb}
\hat{\mathcal{H}}(\Delta v)=\hat{\mathcal{T}}+\hat{U}+\dfrac{\Delta
v}{2}(\hat{n}_1-\hat{n}_0),
\end{eqnarray}
where
\begin{eqnarray}\label{eq:diffpot}
\Delta v=v_1-v_0.
\end{eqnarray}
 In this work, we will consider the singlet two-electron ground and
first excited states for which analytical solutions exist (see
Refs.~\cite{carrascal2015hubbard,PRB16_Burke_thermal_DFT_dimer} and the Appendix). 
Note that, in order to yield the first singlet transition, the minimization in the GOK variational principle (see
Eq.~(\ref{eq:GOK_VP_Tr})) can be restricted to singlet
wavefunctions, since singlet and triplet states are not coupled.
Consequently, eDFT can be formulated for singlet ensembles only. Obviously,
in He for example, singlet eDFT would not describe the lowest 
transition $1^1S\rightarrow 2^3S$. In the following, the first singlet excited
state (which is the excited state studied in this work) will be referred
to as "first excited state" for simplicity.\\

For convenience, the occupation of site
0 is denoted $n_0=n$ and we have $n_1=2-n$ since the number of electrons is held
constant and equal to 2. Therefore, in this simple system, the density
is given by a single number $n$ that can vary from 0 to 2. Consequently, in
this context, DFT becomes a site-occupation functional
theory~\cite{chayes1985density,gunnarsson1986density,PRB95_Gunnarsson_soft,DFT_ModelHamiltonians}
and the various functionals introduced previously will now be functions
of $n$. The ensemble LL functional in Eq.~(\ref{eq:general_LL_ensfun}) becomes
\begin{eqnarray}\label{eq:Hubb_general_LL_ensfun}
F^{w}(n)&=&\underset{\hat{\gamma}^{{w}}\rightarrow n}{\rm
min}\left\{{\rm Tr}\left[\hat{\gamma}^{{w}}(\hat{\mathcal{T}}+\hat{U}
)\right]\right\},
\end{eqnarray}
where the density constraint reads ${\rm
Tr}\left[\hat{\gamma}^{{w}}\hat{n}_0\right]=n$. By analogy with 
Eq.~(\ref{eq:legendre_fenchel_ensembleLL}) and using $n_1-n_0=2(1-n)$, we obtain the following Legendre--Fenchel transform expression, 
\begin{eqnarray}\label{eq:LF_Hubb}
F^{w}(n) =  \sup_{\Delta v} \Big\{
&&(1-w)
E_{0}(\Delta v) + w E_{1}(\Delta v)
\nonumber\\ 
&&+ \Delta v\times (n-1) \Big\},
\end{eqnarray}
where $E_{0}(\Delta v)$ and $E_{1}(\Delta v)$ are the ground- and
first-excited-state energies of $\hat{\mathcal{H}}(\Delta v)$. Note that, even
though analytical expressions exist for the energies, $F^{w}(n)$ has no
simple expression in terms of the density $n$. Nevertheless, as readily
seen in Eq.~(\ref{eq:LF_Hubb}), it can be computed exactly by performing
so-called Lieb maximizations. Note that an accurate parameterization has
been provided by Carrascal~\etal~\cite{carrascal2015hubbard} for the
ground-state LL functional ($w=0$).\\

Similarly, the ensemble non-interacting kinetic
energy in Eq.~(\ref{eq:legendre_fenchel_ensembleTs}) becomes
\begin{eqnarray}\label{eq:Hubb_Tsw}
T_s^{w}(n) =  \sup_{\Delta v} \Big\{
&&(1-w)
\mathcal{E}_0^{KS}(\Delta v) + w \mathcal{E}_1^{KS}(\Delta v)
\nonumber\\ 
&&+ \Delta v \times(n-1) \Big\},
\end{eqnarray}
where $\mathcal{E}_0^{KS}(\Delta v)$ and $\mathcal{E}_1^{KS}(\Delta
v)$ are the ground- and first-excited-state energies of the KS
Hamiltonian 
\begin{eqnarray}\label{eq:KS_hamil_dimer}
\hat{\mathcal{H}}^{KS}\left(\Delta v\right)=\hat{\mathcal{T}}+\frac{\Delta v}{2}(\hat{n}_1-\hat{n}_0).
\end{eqnarray}
From the simple
analytical expressions for the HOMO
and LUMO energies,
\begin{eqnarray}\label{eq:Hubb_noint_E0}
\varepsilon_H(\Delta v)=-\sqrt{t^2+(\Delta v^2/4)},
\end{eqnarray}
and 
\begin{eqnarray}\label{eq:eLUMO_Hubb}
\varepsilon_L(\Delta v)=-\varepsilon_H(\Delta v),
\end{eqnarray}
it comes that
\begin{eqnarray}\mathcal{E}_0^{KS}(\Delta v)=-2\sqrt{t^2+(\Delta
v^2/4)},
\end{eqnarray}
and
\begin{eqnarray}\label{eq:Hubb_noint_E1}
\mathcal{E}_1^{KS}(\Delta v)=0
.
\end{eqnarray} 
According to the Hellmann--Feynman
theorem, combining Eqs.~(\ref{eq:KS_hamil_dimer}) and (\ref{eq:Hubb_noint_E1}) 
leads to 
\begin{eqnarray}\label{eq:HFth_excited_dens}
&&\dfrac{\partial \mathcal{E}_1^{KS}(\Delta v)}{\partial \Delta
v}=\dfrac{1}{2}
\left\langle \Phi_1^{KS}(\Delta
v)\middle\vert\hat{n}_1-\hat{n}_0\middle\vert \Phi_1^{KS}(\Delta v)\right\rangle
\nonumber\\
&&=1-
\left\langle \Phi_1^{KS}(\Delta
v)\middle\vert\hat{n}_0\middle\vert \Phi_1^{KS}(\Delta v)\right\rangle
=0,
\end{eqnarray}
where $\Phi_1^{KS}(\Delta v)$ is the first singlet (two-electron) excited state
of $\hat{\mathcal{H}}^{KS}\left(\Delta v\right)$. Therefore,    
the density ({\it
i.e.} the occupation of site 0) in the non-interacting first excited
state is equal to 1 for any $t$ and $\Delta v$ values, as illustrated in
the top left-hand panel of Fig.~\ref{fig:densetat}. Consequently, a
density $n$ will be ensemble non-interacting representable in this
context if it can be written as $n=(1-w)n^0+w$ where the non-interacting
ground-state density $n^0$ varies in the range $0 \leq n^0 \leq2$ (see
the top left-hand panel of Fig.~\ref{fig:densetat}),
thus leading to the non-interacting representability condition 
\begin{eqnarray}\label{eq:cond_ens_dens}
w \leq n \leq 2-w,
\end{eqnarray}
or, equivalently,
\begin{eqnarray}\label{eq:cond_ens_dens_absval}
\vert n-1\vert \leq 1-w.
\end{eqnarray}
For such densities,    
the maximizing KS potential in Eq.~(\ref{eq:Hubb_Tsw})
equals
\begin{eqnarray}\label{eq:vksw_analytical}
 \Delta v^{KS,w}(n) = \frac{2 (n-1) t}{\sqrt{(1-w)^{2} - (1-n)^{2}}},\label{dvks}
 \end{eqnarray}
and, consequently, the ensemble non-interacting kinetic energy
functional can be expressed analytically as follows, 
\begin{eqnarray}\label{eq:Tsw}
T_s^{w}(n)=-2t\sqrt{(1-w)^{2} - (1-n)^{2}}.
\end{eqnarray}
The ensemble correlation energy, which is the key quantity studied in
this work, is defined as follows,
\begin{eqnarray}\label{eq:ens_corr_energy}
E^{w}_{c}(n)=F^{w}(n)-T_s^{w}(n)-E_H(n)-E^{w}_{x}(n),
\end{eqnarray}
where the Hartree energy
equals~\cite{PRB16_Burke_thermal_DFT_dimer}  
\begin{eqnarray}\label{eq:hartree_fun_def_hubbard_dimer}
E_H(n)&=&
\dfrac{U}{2}\Big(n^2_0+n_1^2\Big)
\nonumber\\
&=&U\Big(1+(1-n)^2\Big).
\end{eqnarray}
Note that the latter expression is simply obtained from the conventional
one in Eq.~(\ref{eq:EH_def}) by substituting a Dirac-delta interaction with strength $U$
for the regular two-electron repulsion, 
\begin{eqnarray}
\dfrac{1}{\vert{\bf r}-{\bf
r}'\vert}\rightarrow U\delta ({\bf r}-{\bf r}'),
\end{eqnarray}
and by summing over sites rather than integrating over the (continuous) real space.  
The exact ensemble exchange energy in Eq.~(\ref{eq:ens_EXX_def}) becomes
in this context 
\begin{eqnarray}\label{eq:Hubb_ens_EXX}
E^{w}_{x}(n) &=& (1-w)\bra{\Phi^{KS,w}_{0}(n)}\hat{U}\ket{\Phi^{KS,w}_{0}(n)}
\nonumber\\
&&+ w\bra{\Phi^{KS,w}_{1}(n)}\hat{U}\ket{\Phi^{KS,w}_{1}(n)}-E_H(n),
 \end{eqnarray}
thus leading, according to the Appendix, to the analytical expression
\begin{eqnarray}\label{eq:Hubb_ens_EXX_ana}
E^{w}_{x}(n)&=&\dfrac{U}{2}\left[1+w-\dfrac{(3w-1)(1-n)^2}{(1-w)^2}\right]-E_H(n),
\nonumber
\\
&=&
E^{w=0}_{x}(n)+\dfrac{Uw}{2}\left[1-\dfrac{(1-n)^2(1+w)}{(1-w)^2}\right],
\end{eqnarray} 
where 
\begin{eqnarray}\label{eq:x_fun_GS}
E^{w=0}_{x}(n)=-E_H(n)/2
\end{eqnarray}
 is the ground-state exchange energy for
two unpolarized electrons. 
Note that the exchange contribution to the GACE integrand (see
Eq.~(\ref{eq:int_GACE_integrand})) will therefore have a simple analytical expression,   
\begin{eqnarray}\label{eq:x_integrand_gace}
\Delta^w_x(n)&=&\dfrac{\partial E^{w}_{x}(n)}{\partial w}
\nonumber
\\
&=&
\dfrac{U}{2}\left[1-\dfrac{(1-n)^2(1+3w)}{(1-w)^3}\right]
.
\end{eqnarray}

Finally, the maximizing potential $\Delta v^w(n)$ in
Eq.~(\ref{eq:LF_Hubb}) which reproduces the ensemble density $n$
fulfills, according to the inverse Legendre--Fenchel transform,
\begin{eqnarray}\label{eq:invLF_Hubb}
&&(1-w)
E_{0}(\Delta v^w(n)) + w E_{1}(\Delta v^w(n))
\nonumber
\\
 &&=  \min_{\nu} \Big\{F^{w}(\nu) 
- \Delta v^w(n)\times (\nu-1) \Big\},
\end{eqnarray}
where the minimizing density is $n$. Therefore,
\begin{eqnarray}\label{eq:physpot_derivFw}
\Delta v^w(n)=\dfrac{\partial F^w(n)}{\partial n},
\end{eqnarray}
and, since 
(see
Eqs.~(\ref{eq:vksw_analytical}) and (\ref{eq:Tsw}))
\begin{eqnarray}\label{eq:deriv_Tsx_pot}
\Delta v^{KS,w}(n)=\partial T_s^{w}(n)/\partial n,
\end{eqnarray}
 the ensemble
Hartree-$xc$ 
potential reads
\begin{eqnarray}\label{eq:ehxc_pot_derivative}
\Delta v_{Hxc}^w(n)&=&\Delta v^{KS,w}(n)-\Delta v^w(n)
\nonumber\\
&=&-\dfrac{\partial
E^w_{Hxc}(n)}{\partial n}.
\end{eqnarray}
As a result, the ensemble correlation potential can be calculated
exactly as
follows,      
\begin{eqnarray}
\Delta v_{c}^w(n)&=&\Delta v^{KS,w}(n)-\Delta v^w(n)
\nonumber\\
&&-\Delta
v_{H}(n)-\Delta v_x^w(n),
\end{eqnarray}
where all contributions but $\Delta v^w(n)$ have an analytical
expression. The Hartree potential equals $\Delta v_{H}(n)=-\partial 
E_H(n)/\partial n=2U(1-n)$ and, according to
Eq.~(\ref{eq:Hubb_ens_EXX_ana}), the ensemble exchange potential
reads
\begin{eqnarray}\label{eq:exact_ens_X_pot}
\Delta v_x^w(n)&=&-\dfrac{\partial E^{w}_{x}(n)}{\partial n}
\nonumber\\
&=&
U(n-1)
\left[1+\dfrac{w(1+w)}{(1-w)^2}\right]
\nonumber\\
&=&\Delta v_x^{w=0}(n)
\left[1+\dfrac{w(1+w)}{(1-w)^2}\right]
.
\end{eqnarray}
Note the unexpected minus sign on the right-hand side of
Eq.~(\ref{eq:ehxc_pot_derivative}). It originates from the
definition of the potential difference (see Eq.~(\ref{eq:diffpot})) and
the choice of $n_0=n$ (occupation of site 0) as variable, the occupation of site 1 being
$n_1=2-n$. Therefore, $E^w_{Hxc}(n)$ can be rewritten as
$E^w_{Hxc}[n,2-n]$
and
\begin{eqnarray}
\Delta v^w_{Hxc}(n)&=&
\left.
\frac{\partial E^w_{Hxc}[n_0,n_1]}{\partial n_1}
\right|_{n_0=n,n_1=2-n}
\nonumber
\\
&&-
\left.
\frac{\partial E^w_{Hxc}[n_0,n_1]}{\partial n_0}
\right|_{n_0=n,n_1=2-n}
\nonumber
\\
&=&
-\frac{\partial E^w_{Hxc}[n,2-n]}{\partial n}=-\frac{\partial
E^w_{Hxc}(n)}{\partial n}.
\end{eqnarray}
Note finally that, as readily seen in Eq.~(\ref{eq:exact_ens_X_pot}),
the ensemble $x$ potential can be expressed in
terms of the ground-state $x$ potential ($w=0$) and the ensemble weight.
This simple relation, which is transferable to {\it ab
initio} Hamiltonians, could be used for developing "true" approximate
weight-dependent density-functional $x$ potentials. 

\section{Exact results}\label{sec:results}

\subsection{Interacting ensemble density and derivative discontinuity}

\begin{figure}
\begin{center}
\includegraphics[width=0.8\textwidth]{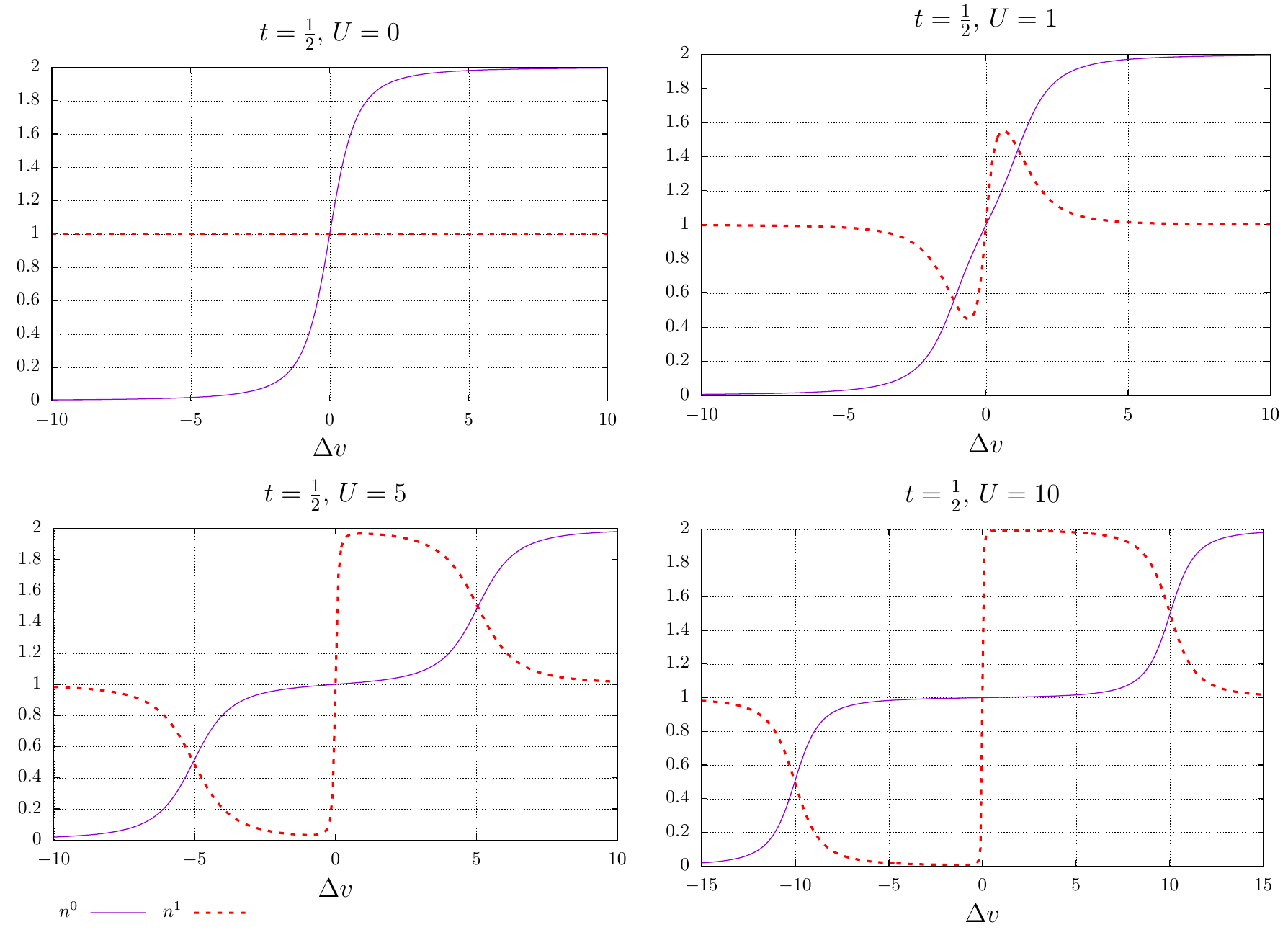}
\end{center}
\caption{Variation of the ground- $n^{0}$ and first-excited-state
$n^{1}$ densities with the local potential $\Delta v$ in
the Hubbard dimer for various $U$ values.}
\label{fig:densetat}
\end{figure}

In the rest of the paper, the hopping parameter is set to $t=1/2$. For clarity, we shall refer to the local potential in the physical
(fully-interacting) Hubbard Hamiltonian as $\Delta v_{ext}$. This potential is the analog of the
nuclear-electron attraction potential in the {\it ab initio} Hamiltonian. The
corresponding ensemble density is the weighted sum of the ground-
$n_{\Delta v_{ext}}^0$ and excited-state $n_{\Delta v_{ext}}^1$ occupations of site 0,
\begin{eqnarray}\label{eq:true_phys_ensdensity0}
n^w=(1-w)n^0_{\Delta v_{ext}}+wn^1_{\Delta v_{ext}},
\end{eqnarray}    
where, according to 
the Hellmann--Feynman theorem,
\begin{eqnarray}\label{eq:true_phys_ensdensityHF}
n_{\Delta v_{ext}}^i=1-\left.\dfrac{\partial E_i(\Delta v)}{\partial \Delta
v}\right|_{\Delta v_{ext}}.
\end{eqnarray}    
Note that the first-order derivative of the energies with respect to
$\Delta v$ can be simply expressed in terms of the
energies (see Eq.~(\ref{eq:deriv_ener_deltav})) and that, for a fixed
$\Delta v_{ext}$ value, the ensemble density varies linearly with $w$. Ground- and excited-state densities are shown in Fig.~\ref{fig:densetat}.
For an arbitrary potential value $\Delta v_{ext}=\Delta v$, in the
weakly correlated regime ($0<U<<\Delta v$), site occupations are close to 2 or
0 in the ground state and they become equal to 1
in the first excited state. Therefore, in this case, the model describes a charge
transfer excitation. On the other hand, in the strongly correlated
regime ($U>>\Delta v$), the ground-state density will be close to 1
(symmetric case). 
When $U$ is large,
small changes in $\Delta v$ around $\Delta v=0$ cause large changes in the 
excited-state density. As clearly seen from the Hamiltonian
expression in Eq.~(\ref{eq:final_ham_Hubb}), when $U\rightarrow+\infty$, site 0 "gains" an
electron when the lowest (singlet) transition occurs if $\Delta
v\rightarrow 0^+$ whereas, if $\Delta v\rightarrow 0^-$, it "loses" an electron. This explains why  
the excited-state density curves approach a discontinuous limit at
$\Delta v=0$ when $U\rightarrow+\infty$. Let us stress that, for large but finite $U$
values, the latter density will vary rapidly and continuously from 0 to
2 in the vicinity of $\Delta v=0$ while the ground-state density remains
close to 1. This observation will enable us to interpret the
GACE integrand in the following.\\               

\begin{figure}
 
\begin{center}
\includegraphics[width=1.0\textwidth]{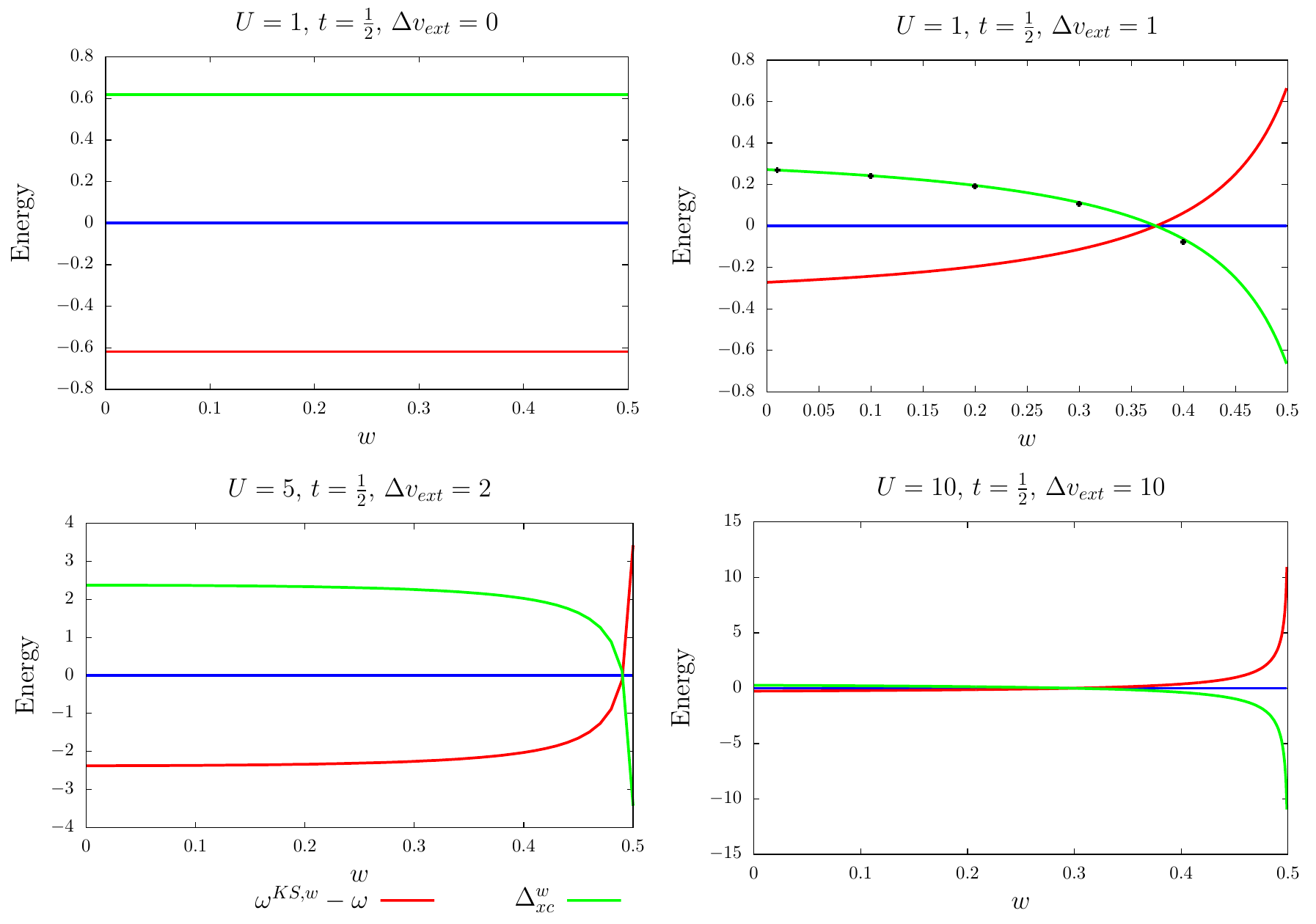}
\end{center}
\caption{Derivative discontinuity obtained for the Hubbard dimer with
different $\Delta v_{ext}$ and $U$ values. Results obtained by numerical
differentiation are shown for $\Delta v_{ext}=1$ and
$U=1$ (see the black dots on the right-hand top
panel). See text for further details.
}
\label{fig:fish1}
\end{figure}

\begin{figure}
\centering
\includegraphics{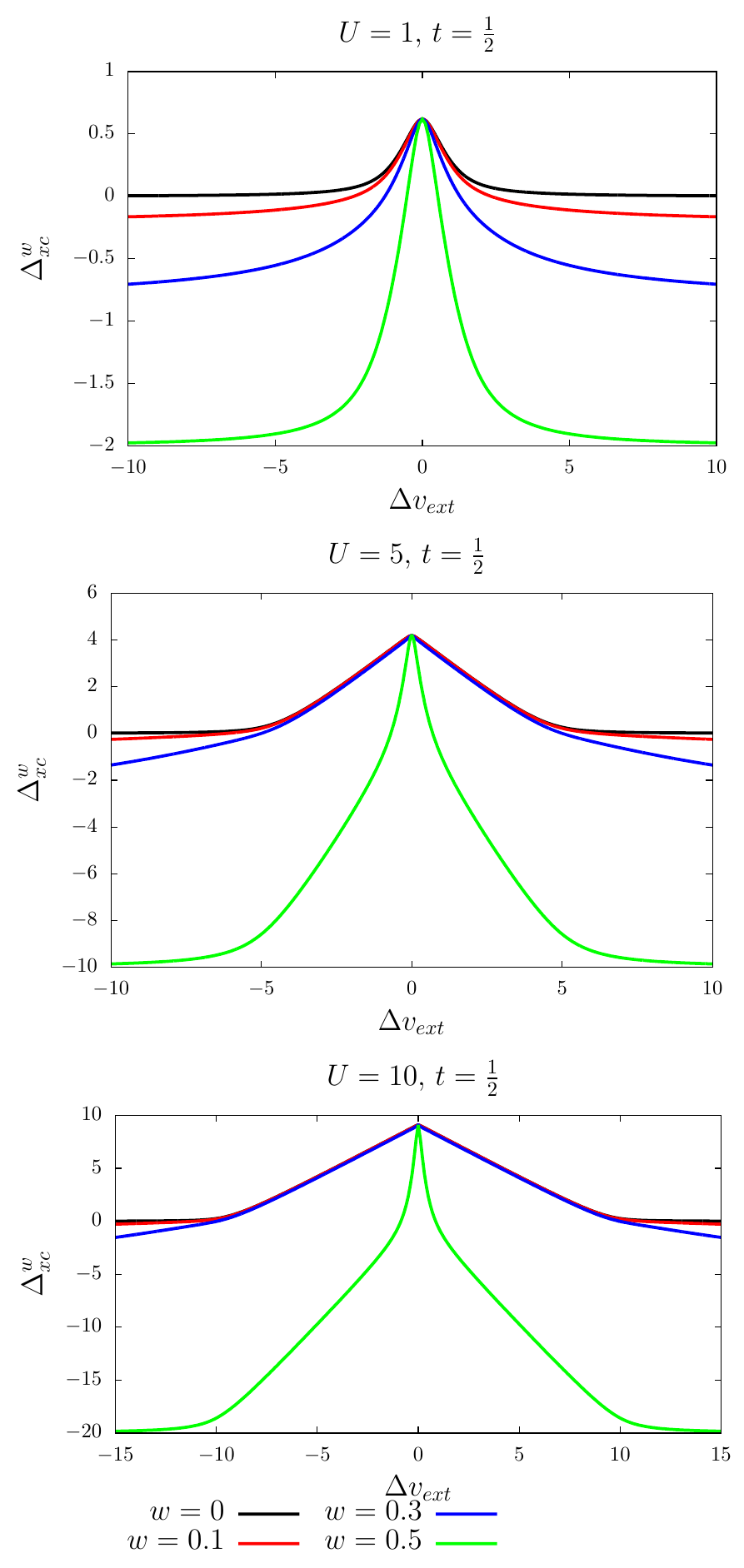}
\caption{Derivative discontinuity plotted as a function of $\Delta
v_{ext}$ for various $w$ and $U$ values.}\label{fig:xcDD_wrt_pot}
\end{figure}

\begin{figure}
\begin{center}
\includegraphics[width=1.0\textwidth]{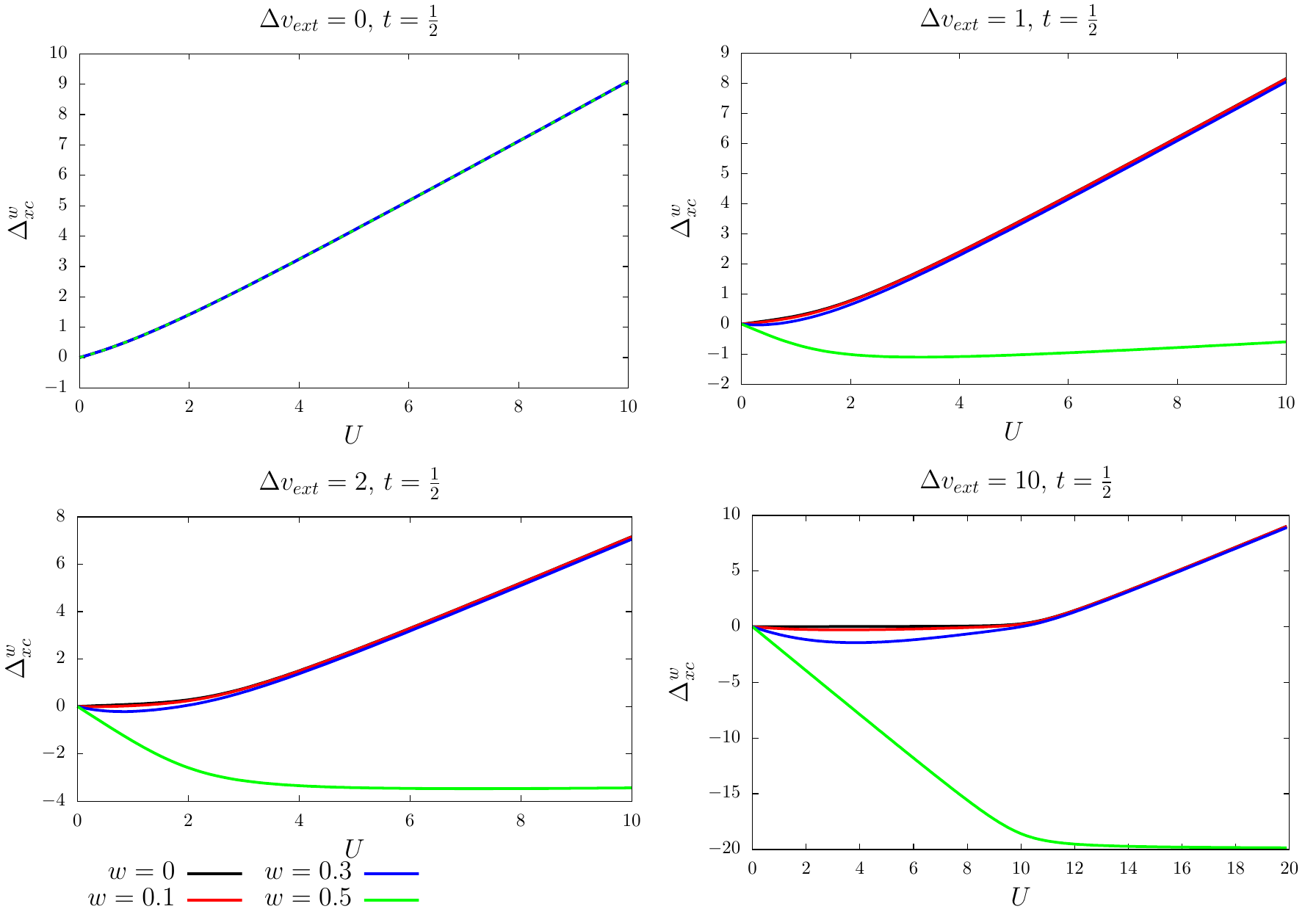}
\caption{Derivative discontinuity plotted as a function of $U$ for
various $w$ and $\Delta v_{ext}$ values.}\label{fig:xcDD_wrt_U}
\end{center}
\end{figure}

Turning to the calculation of the DD (see
Eq.~(\ref{eq:xcDD})), the latter can be obtained in two ways, either by
taking the difference between the physical $\omega=E_1(\Delta
v_{ext})-E_0(\Delta v_{ext})$ and KS
\begin{eqnarray}
\omega^{KS,w}=\varepsilon_L\Big(\Delta
v^{KS,w}(n^w)\Big)-\varepsilon_H\Big(\Delta v^{KS,w}(n^w)\Big)
\end{eqnarray}
excitation energies, which gives  
\begin{eqnarray}
\Delta^w_{xc}
&=&\omega-\omega^{KS,w}, 
\end{eqnarray}  
or by differentiation, 
\begin{eqnarray}\label{eq:DD_finite_diff}
\Delta^w_{xc}&=&\left.\dfrac{\partial E^\xi_{xc}(n^w)}{\partial
\xi}\right|_{\xi=w}.
\end{eqnarray}  
In the former case, we obtain from Eqs.~(\ref{eq:Hubb_noint_E0}),
(\ref{eq:eLUMO_Hubb}), and (\ref{eq:vksw_analytical}) the analytical expression
\begin{eqnarray}\label{eq:DD_exp_Hubb}
\Delta^w_{xc}
&=&E_1(\Delta v_{ext})-E_0(\Delta v_{ext})
\nonumber
\\
&&-\dfrac{2t(1-w)}{\sqrt{(1-w)^2-(1-n^w)^2}}.
\end{eqnarray}  
Regarding Eq.~(\ref{eq:DD_finite_diff}), the $\xi$-dependent ensemble
$xc$ 
energy $E^\xi_{xc}(n^w)$ must be
determined numerically by means of a Legendre--Fenchel transform
calculation (see Eqs.~(\ref{eq:LF_Hubb}) and (\ref{eq:ens_corr_energy})) and its derivative at $\xi=w$ is then obtained by finite
difference. As illustrated in the right-hand top panel of
Fig.~\ref{fig:fish1}, the two expressions are indeed
equivalent. In the symmetric Hubbard dimer ($\Delta v_{ext}=0$), it is clear from
Eq.~(\ref{eq:DD_exp_Hubb}) that the DD is weight-independent, since $n^w=1$, and it is equal to 
$[U-4t+\sqrt{U^2+16t^2}]/2$. In this particular
case, the ground and first-excited states actually belong to different
symmetries. In the asymmetric case, various patterns are obtained (see
Fig.~\ref{fig:fish1}). Interestingly, the "fish picture" obtained by  
Yang~\etal~\cite{PRA14_Burke_exact_GOK-DFT} for the helium atom is
qualitatively reproduced by the Hubbard dimer model when $\Delta
v_{ext}=U=1$, except in the small-$w$ region where a sharp change in the
DD (with
positive slope) is observed for the helium atom. 
This feature does not occur in the two-site model.
From the analytical expression, 
\begin{eqnarray}
\dfrac{\partial \Delta^w_{xc}}{\partial
w}=\dfrac{2t(1-n^w)(n^1-1)}{\left[(1-w)^2-(1-n^w)^2\right]^{3/2}},
\end{eqnarray} 
and Fig.~\ref{fig:densetat}, it becomes clear that, in the Hubbard
dimer, the DD will systematically decrease with
$w$. Variations in $\Delta v_{ext}$ and $U$ for various weights are
shown in Figs.~\ref{fig:xcDD_wrt_pot} and \ref{fig:xcDD_wrt_U},
respectively.
When $\Delta v_{ext}>>U$, $n^w$ is close to $2-w$
(according to Fig.~\ref{fig:densetat}) and, since the on-site repulsion
becomes a perturbation, the DD can be well reproduced by the
exchange-only contribution. Thus,
according to Eq.~(\ref{eq:x_integrand_gace}), we obtain
\begin{eqnarray}\label{eq:DD_vext_large}
\Delta^w_{xc}\rightarrow \Delta^w_{x}(n^w)\approx
-\dfrac{2Uw}{(1-w)}
.
\end{eqnarray}
As readily seen in Eq.~(\ref{eq:DD_vext_large}), the DD is close to zero
for small weights and, when $w=1/2$, it equals $-2U$, which is in agreement with
both Figs.~\ref{fig:xcDD_wrt_pot} and \ref{fig:xcDD_wrt_U}. On the other
hand, when $t<<\Delta v_{ext}<<U$, the physical energies are expanded as
follows, according to Eq.~(\ref{eq:3rdorder_energy}),   
\begin{eqnarray}\label{eq:exp_indiv_ener_toverU}
E_0(\Delta v_{ext})/U&=&\dfrac{4}{(\Delta
v_{ext}/U)^2-1}(t/U)^2+\mathcal{O}\left((t/U)^3\right)
\nonumber\\
E_1(\Delta v_{ext})/U&=&1-(\Delta v_{ext}/U)
+\dfrac{2}{1-(\Delta
v_{ext}/U)}(t/U)^2
\nonumber
\\
&&
+\mathcal{O}\left((t/U)^3\right),
\end{eqnarray}
thus leading to the following expansions for the derivatives,  
\begin{eqnarray}
\dfrac{\partial E_0(\Delta v_{ext})}{\partial \Delta
v_{ext}}&=&-\dfrac{8(\Delta v_{ext}/U)}{\left[(\Delta
v_{ext}/U)^2-1\right]^2}(t/U)^2+\mathcal{O}\left((t/U)^3\right)
\nonumber\\
\dfrac{\partial E_1(\Delta v_{ext})}{\partial \Delta v_{ext}}&=&-1
+\dfrac{2}{\left[1-(\Delta
v_{ext}/U)\right]^2}(t/U)^2
\nonumber
\\
&&
+\mathcal{O}\left((t/U)^3\right),
\end{eqnarray}
and, according to Eqs.~(\ref{eq:true_phys_ensdensity0}) and
(\ref{eq:true_phys_ensdensityHF}), to the following expansion for the
ensemble density,
\begin{eqnarray}\label{eq:ensdens_expans_toverU}
&&n^w=1+w
\nonumber\\
&&+
\dfrac{2(t/U)^2}{\left[1-(\Delta v_{ext}/U)\right]^2}
\Bigg[\dfrac{4(1-w)(\Delta v_{ext}/U)}{\left[1+(\Delta
v_{ext}/U)\right]^2}-w\Bigg]
\nonumber\\
&&+\mathcal{O}\left((t/U)^3\right).
\end{eqnarray}
As readily seen in Eq.~(\ref{eq:ensdens_expans_toverU}), the ensemble
density is close to 1 in the small-$w$ region. Consequently, according
to Eqs.~(\ref{eq:DD_exp_Hubb}) and (\ref{eq:exp_indiv_ener_toverU}),  
the DD varies as $U-\Delta v_{ext}$, which is in
agreement with the $U=10$ panel of Fig.~\ref{fig:xcDD_wrt_pot} and the $\Delta v_{ext}=10$ panel of Fig.~\ref{fig:xcDD_wrt_U}. On the other
hand, when $w=1/2$, it comes from Eq.~(\ref{eq:ensdens_expans_toverU}),  
\begin{eqnarray}\label{eq:ensdens_expans_toverU_wonehalf}
\dfrac{1}{4}-\big(1-n^{w=1/2}\big)^2&=&
\dfrac{(t/U)^2}{\left[1+(\Delta v_{ext}/U)\right]^2}
\nonumber\\
&&
+\mathcal{O}\left((t/U)^3\right),
\end{eqnarray}
thus leading to the following expansion for the equiensemble DD,
\begin{eqnarray}  
\Delta^{w=1/2}_{xc}/U&=&-2(\Delta v_{ext}/U)
+\mathcal{O}(t/U).
\end{eqnarray}
The latter expansion matches the behavior observed in the $U=5$ and $U=10$ panels of
Fig.~\ref{fig:xcDD_wrt_pot} as well as $\Delta v_{ext}=2$ and $\Delta
v_{ext}=10$ panels of Fig.~\ref{fig:xcDD_wrt_U}, when $t<<\Delta v_{ext}<<U$. Note finally that, in
the $U=10$ panel of Fig.~\ref{fig:xcDD_wrt_pot}, the equiensemble DD is
highly sensitive to changes in $\Delta v_{ext}$ around $\Delta
v_{ext}=0$ when $U>>t$. In the latter case, the ground-state density remains
close to 1 (symmetric dimer), as shown in Fig.~\ref{fig:densetat}, and the DD becomes
\begin{eqnarray}\label{eq:DD_close_to_symm}
\Delta^w_{xc}&\rightarrow& \dfrac{1}{2}\left[U+\sqrt{U^2+16t^2}\right]
\nonumber\\
&&-\dfrac{2t(1-w)}{\sqrt{
1-2w+n^1_{\Delta v_{ext}}(2-n^1_{\Delta v_{ext}})w^2
}},
\end{eqnarray}
 which is almost constant in the small-$w$ region. When $w=1/2$, the
second term on the right-hand side of Eq.~(\ref{eq:DD_close_to_symm})
becomes $-2t/\sqrt{n^1_{\Delta v_{ext}}(2-n^1_{\Delta v_{ext}})}$, which
decreases rapidly with $\Delta v_{ext}$ as the excited-state density
approaches (also rapidly) 2.\\ 

\begin{figure}
\centering
\includegraphics{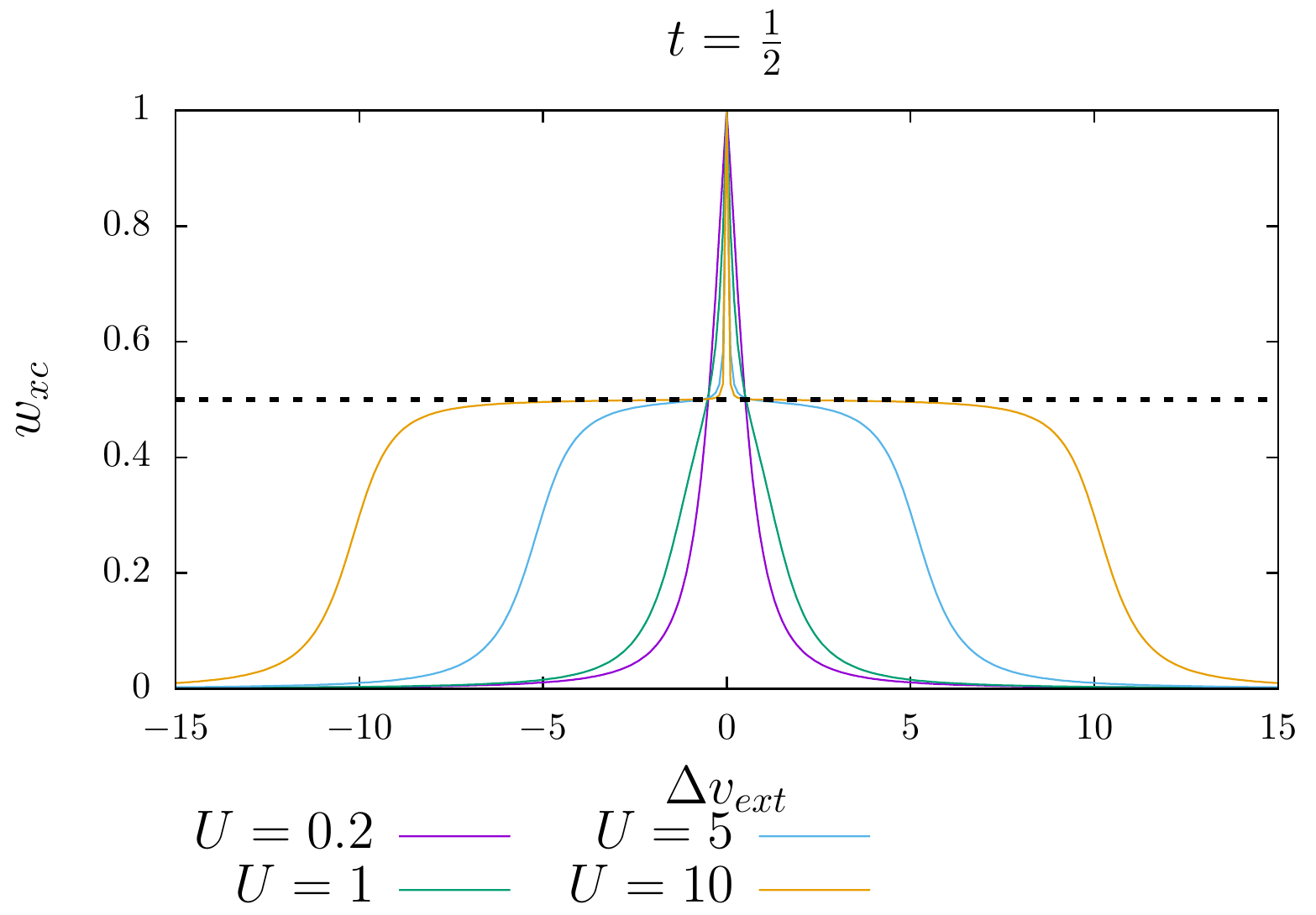}
\caption{
Solution of Eq.~(\ref{eq:2ndorder_polyn_wxc}) plotted with respect to the potential $\Delta
v_{ext}$ for different $U$ values. Physical values should be lower than
1/2 ({\it i.e.} below the dashed line).}
\label{fig:w0}
\end{figure}

Let us finally focus on the weight $w_{xc}$ for which the DD
vanishes:
\begin{eqnarray}\label{eq:w_xc_def}
\Delta^{w_{xc}}_{xc}
=\displaystyle\left.\frac{\partial E^{w
}_{xc}(n^{w_{xc}})}{\partial
w}\right|_{
w=w_{xc}}=0.
\end{eqnarray}
For that particular weight, which should of course be used in both KS
and physical systems, the (weight-dependent) KS HOMO-LUMO gap
is equal to the exact physical (weight-independent) excitation energy,
which is remarkable. Note that $w_{xc}$, if it exists, would be fully
determined, in practice, from the "universal" ensemble $xc$
functional. Indeed, for a given local potential $\Delta v_{ext}$, the ensemble density $n^w$ (see
Eq.~(\ref{eq:true_phys_ensdensity0})) can be obtained by solving two
self-consistent KS equations. 
One with $w=0$ (which gives the ground-state
density $n_{\Delta v_{ext}}^0$) and a second one with $w=1/2$. In the latter case,
\begin{eqnarray}
n^{w=1/2}=(n_{\Delta v_{ext}}^0+n_{\Delta v_{ext}}^1)/2,
\end{eqnarray}
thus leading to $n_{\Delta v_{ext}}^1=2n^{w=1/2}-n_{\Delta v_{ext}}^0$.
The value of $w_{xc}$ would then be obtained from Eq.~(\ref{eq:w_xc_def}). Solving the ensemble KS equations
with the weight $w_{xc}$ would lead to a KS gap which is, in this particular
case, the physical optical one. Note that, even though the DD equals zero in this case, it is necessary to know the weight dependence of the
ensemble $xc$ functional in order to determine $w_{xc}$. Despite the
simplicity of the Hubbard dimer model, $E^w_{xc}(n)$ cannot (like in the
ground-state case~\cite{carrascal2015hubbard}) be expressed
analytically in terms of $n$ and $w$. The exact
value of $w_{xc}$ has been simply determined from
Eq.~(\ref{eq:DD_exp_Hubb}), where the exact physical excitation energy
$\omega$
is known, thus leading to the second-order polynomial equation,
\begin{eqnarray}\label{eq:2ndorder_polyn_wxc}
&&w_{xc}^2\Big[
\omega^2
-\omega^2\left(n_{\Delta v_{ext}}^1-n_{\Delta
v_{ext}}^0\right)^2
-4t^2\Big]
\nonumber\\
&&+2w_{xc}\Big[
\omega^2\left(n_{\Delta v_{ext}}^0-n_{\Delta
v_{ext}}^1\right)
\left(n_{\Delta v_{ext}}^0-1\right)
\nonumber\\
&&-\omega^2+4t^2\Big]
+\omega^2n_{\Delta v_{ext}}^0(2-n_{\Delta v_{ext}}^0)-4t^2=0.
\end{eqnarray}
Physical solutions should be in the range $0\leq w_{xc}\leq 1/2$.
Results are shown in 
Fig.~\ref{fig:w0}. 
In the symmetric Hubbard dimer, the solution becomes $w_{xc}=1$, which is
unphysical. This is in agreement with the fact that, in this case, the
DD is constant and strictly positive. This is also the reason why no physical values are obtained for $w_{xc}$ in
the vicinity of $\Delta v_{ext}=0$. Note finally that 
$w_{xc}$ is quite sensitive to changes in $\Delta
v_{ext}$ around $\Delta v_{ext}=U$ in both weak and strong correlation
regimes.
This indicates that 
$w_{xc}$ strongly depends on the system under study.   

\subsection{Construction and analysis of the
GACE}\label{subsec:GACE_analysis}

\begin{figure}

\begin{center}
\includegraphics[width=1.0\textwidth]{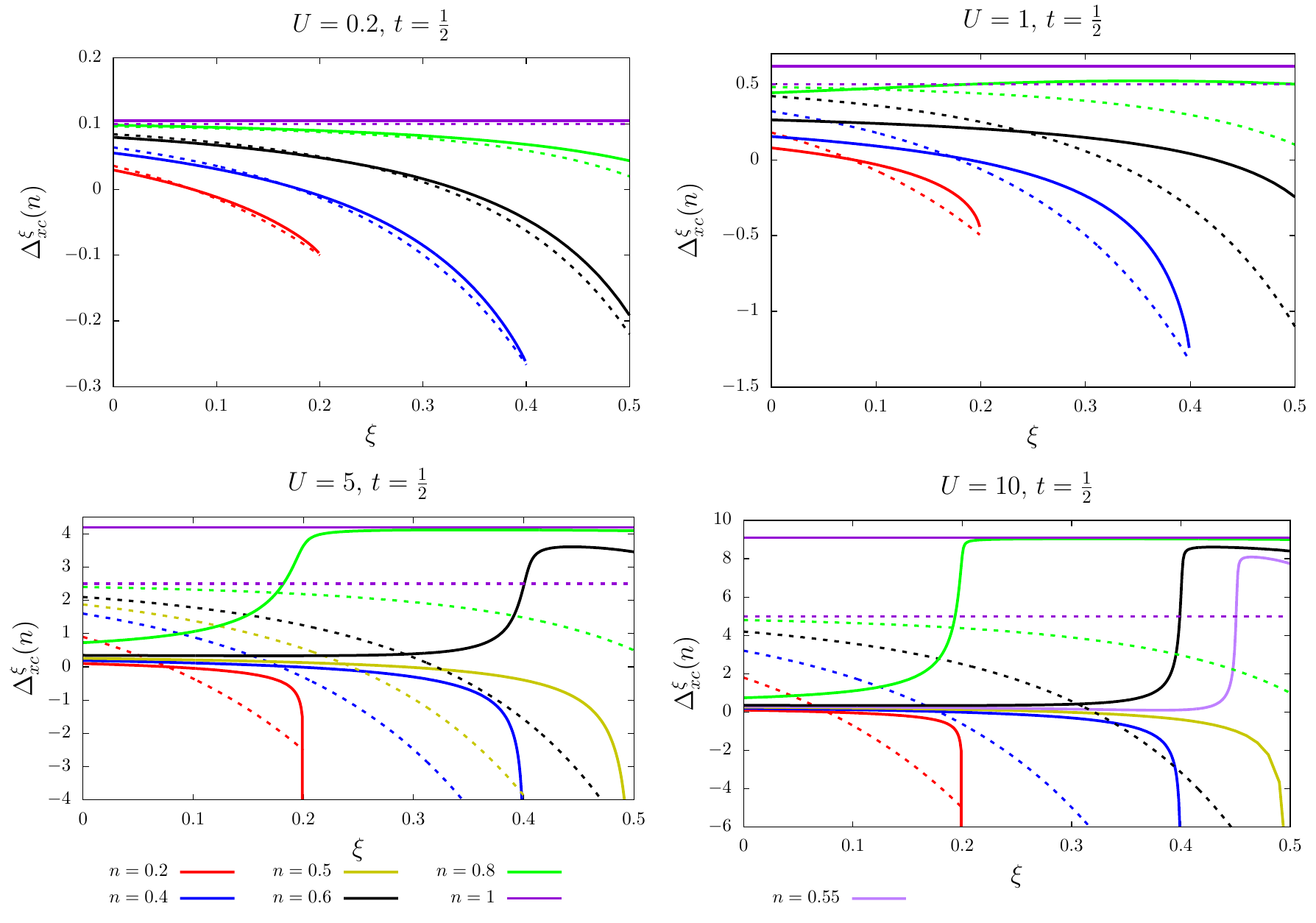}
\end{center}
\caption{Variation of the exchange-correlation GACE integrand with
the ensemble weight $\xi$ for
various $U$ values and densities. Comparison is made with the exact exchange-only contribution
$\Delta^{\xi}_{x}(n)$ (dashed lines).}
\label{fig:gace_integrand}
\end{figure}

The general GACE integrand expression in
Eq.~(\ref{eq:general_gace_integrand}) can, in
the case of the Hubbard dimer,
be simplified as follows,
\begin{eqnarray}\label{eq:DD_fun}
\Delta^\xi_{xc}(n)
&=&E_1\Big(\Delta v^\xi(n)\Big)-E_0\Big(\Delta v^\xi(n)\Big)
\nonumber
\\
&&-\dfrac{2t(1-\xi)}{\sqrt{(1-\xi)^2-(1-n)^2}},
\end{eqnarray}  
where the local potential $\Delta v^\xi(n)$ can be computed exactly 
by means of the
Legendre--Fenchel transform in Eq.~(\ref{eq:LF_Hubb}). 
Results are shown in Fig.~\ref{fig:gace_integrand}.
Note that, for a fixed density $n$, the non-interacting
$v$-representability condition for an ensemble weight $\xi$ (see
Eq.~(\ref{eq:cond_ens_dens})) reads 
\begin{eqnarray}\label{eq:rep_condition}
0\leq\xi \leq 
1-\vert n-1\vert
.
\end{eqnarray} 
In the symmetric case ($n=1$), the weight-independent
value $[U-4t+\sqrt{U^2+16t^2}]/2$ is recovered. 
In the weakly correlated regime ($U=0.2$), the analytical exact exchange
expression for the GACE integrand (see Eq.~(\ref{eq:x_integrand_gace}))
reproduces very well the total $xc$ one, as expected. When $0\leq n \leq
0.5$, the integrand at $\xi=n$ is therefore well approximated by
$\Delta^{\xi=n}_{x}(n)=2Un/(n-1)$. 
Note also that,
away from the symmetric case, the
exchange integrand curve crosses over the $xc$ one so that, after
integration over the ensemble weight, the ensemble correlation energy
remains negligible. In other words, integrals of the exchange and $xc$
integrands are expected to be very similar ({\it i.e.} second order in
$U$), which explains why the curves have to cross when, in the
large-$\xi$ region, the two
integrands differ substantially.  
\\

Let us now focus
on the stronger correlation regimes. For the large $U=5$ and $U=10$
values, we can see plateaus for the
considered $n=0.6$ and $n=0.8$ densities in the range
$1-n\leq \xi\leq1/2$, thus leading to 
discontinuities in the GACE integrand when $U/t\rightarrow+\infty$. As
readily seen in Eq.~(\ref{eq:DD_fun}), these discontinuities are induced by the
$\xi$-dependent fully interacting excitation energy (first term on the
right-hand side). As illustrated in Fig.~\ref{fig:densetat}, when $U$ is large, the density of
the ground state is close to 1 in the vicinity of the symmetric
potential ($\Delta v=0$) while the density of the excited state is
highly sensitive to small changes in the potential. The reason
is that, in the $U/t\rightarrow+\infty$ limit, states with a
doubly-occupied site are degenerate (with energy $U$) when $\Delta
v=0$. The
degeneracy is lifted when $\Delta v$ is not strictly zero. For finite
but large $U/t$ values, the first-excited state density will vary
continuously and rapidly from 0 to 2 in the vicinity of $\Delta v=0$. Therefore, within the GACE, the fully-interacting
ensemble density reads $n=(1-\xi)+\xi n^{1,\xi}$ with the
condition $0\leq n^{1,\xi} \leq2$, thus leading to
\begin{eqnarray}
n^{1,\xi}=1+\dfrac{n-1}{\xi},
\end{eqnarray}              
and $\vert 1-n\vert\leq \xi \leq 1/2$. The latter range describes 
exactly the plateaus observed in the $U=10$ panel of
Fig.~\ref{fig:gace_integrand}. In this case, the GACE potential in the physical system is almost
symmetric, thus leading to the following approximate value for the
plateau,
\begin{eqnarray}\label{eq:DD_plateau_exp}
\Delta^\xi_{xc}(n)&\approx& \dfrac{1}{2}(U+\sqrt{U^2+16t^2})
\nonumber
\\
&&-\dfrac{2t(1-\xi)}{\sqrt{(1-\xi)^2-(1-n)^2}}.
\end{eqnarray}
This expression will be used in the following section for analyzing the
ensemble $xc$ energy and potential. Note that the $\xi$-dependent part
of the integrand (second term on the right-hand side of
Eq.~(\ref{eq:DD_plateau_exp})) decreases with $\xi$ over the range
$(1-n)\leq \xi\leq 1/2$ with $1/2\leq n \leq 1$, as clearly seen in the
$U=5$ and $U=10$ panels of Fig.~\ref{fig:gace_integrand}. The
$\xi$-dependence disappears as $U/t$ increases.\\
   
We also in Fig.~\ref{fig:gace_integrand} that, outside the
plateaus, the GACE integrand becomes relatively small as $U$ increases.   
This can be interpreted as follows. In the $U/t\rightarrow+\infty$
limit, when $\Delta v=\pm U$, the ground (with singly occupied
sites) and first-excited (with a doubly occupied site) states
become degenerate with energy 0. If we consider, for example, an
infinitesimal positive deviation from $-U$ in the potential,
sites will be singly occupied in the ground state and site 0 will be
empty in the first excited state. It would be the opposite if the
deviation were negative, thus leading to discontinuites in the ground-
and excited-state densities at $\Delta v
=\pm U$, as expected from the $U=10$ panel of Fig.~\ref{fig:densetat}.
For large but finite $U/t$ values, the ground-state density will vary
continuously from 0 to 1 around $\Delta v=-U$ 
while the first-excited-state density varies from 1 to 0.
The first excitation
is a charge transfer. It means that, in this case, the
fully-interacting ensemble density with weight $\xi$ can be written as
$n=(1-\xi)n^{0,\xi}+\xi n^{1,\xi}$ with $n^{0,\xi}+n^{1,\xi}=1$, thus
leading to
\begin{eqnarray}
n^{0,\xi}-1=\dfrac{n-1+\xi}{1-2\xi}.
\end{eqnarray}            
Therefore, for a given density $n$, the condition $0\leq n^{0,\xi} \leq 1$ can be 
rewritten as $\xi \leq 1-n$ in addition to the non-interacting
$v$-representability condition in Eq.~(\ref{eq:rep_condition}). Note
that, around $\Delta v=U$, this condition becomes $0\leq \xi \leq 
n-1$. In summary, for a fixed density $n$, the range of ensemble weights
$0\leq \xi\leq \vert 1-n\vert$ can be described
in the vicinity of $\Delta v=\pm U$.
This range corresponds to situations where no plateau is
observed in the GACE integrand. Since, according to
Eq.~(\ref{eq:3rdorder_energy}), the ground- and first-excited-state
energies at $\Delta v=\pm U$ can be expanded as follows,
\begin{eqnarray}
E_0(\pm U)/U&=&-\sqrt{2}t/U+\mathcal{O}(t^2/U^2),
\nonumber
\\
E_1(\pm U)/U&=&\sqrt{2}t/U+\mathcal{O}(t^2/U^2),
\end{eqnarray}
we conclude that, when $0\leq \xi\leq \vert 1-n\vert$ and $U$ is large,
an approximate GACE
integrand expression is
\begin{eqnarray}\label{eq:DD_fun_approx_outside_plateau}
\Delta^\xi_{xc}(n)
&\approx& 2\sqrt{2}t
-\dfrac{2t(1-\xi)}{\sqrt{(1-\xi)^2-(1-n)^2}}.
\end{eqnarray}  
Note that, for an ensemble non-interacting representable density $n$ such
that $n<1$, the condition $\xi\leq n$ must be fulfilled, according to
Eq.~(\ref{eq:rep_condition}). If, in
addition, $n\leq 1-n$ ({\it i.e.} $n\leq 1/2$), then the GACE integrand is 
expected to diverge in the strongly correlated limit when $\xi\rightarrow n$, which      
is exactly what is observed in the $U=10$ panel of
Fig.~\ref{fig:gace_integrand}. 

\subsection{Weight-dependent exchange-correlation energy and potential}

\begin{figure}
\centering
\includegraphics[width=0.6\textwidth]{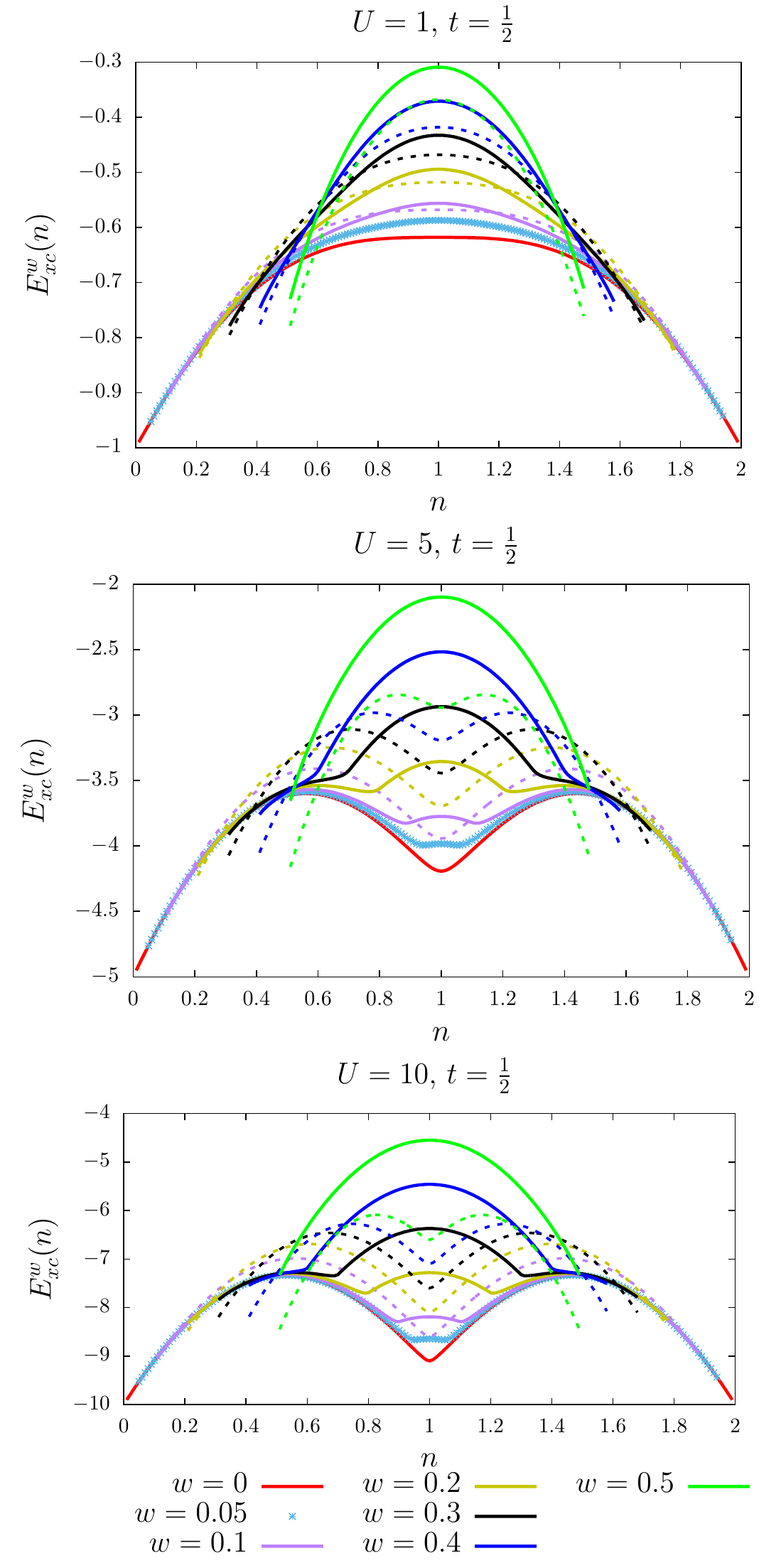}
\caption{Exact ensemble density-functional exchange-correlation energies
for various $U$ and $w$ values. The mixed ensemble exchange/
ground-state correlation energy
$E_x^{w}(n)+E_c^{w=0}(n)$ is shown (dashed lines) for analysis purposes.}\label{fig:eXC_energies}
\end{figure}

\begin{figure}
 
\begin{center}
\includegraphics[width=1.0\textwidth]{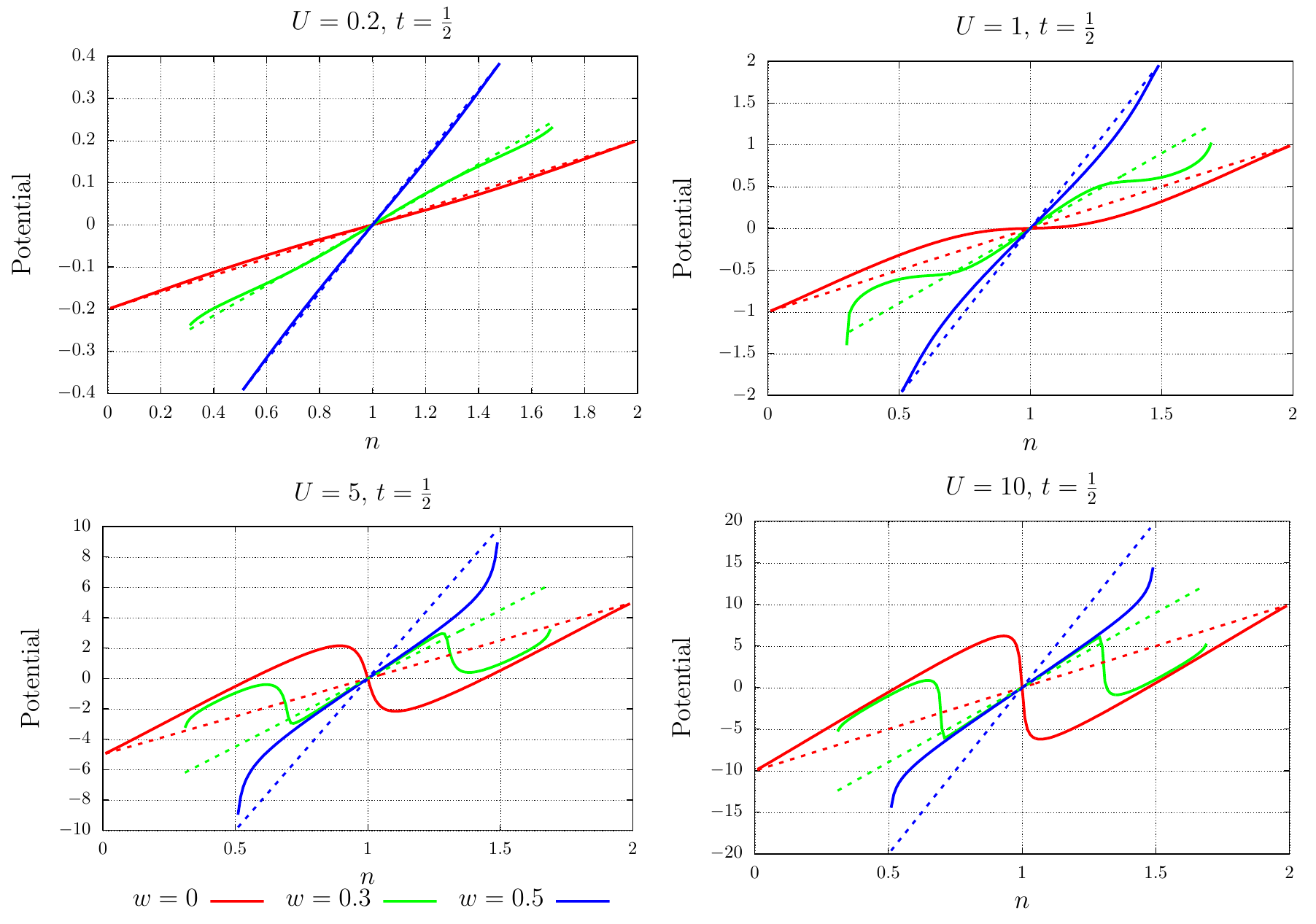}
\end{center}
\caption{Exact ensemble exchange-correlation potential for various $U$
and $w$ values. The exact ensemble exchange potential (dashed lines) is
shown for analysis purposes.}
\label{fig:ens_xc_pot}
\end{figure}

Exact ensemble $xc$ density-functional energies are shown in
Fig.~\ref{fig:eXC_energies}. 
As discussed just after Eq.~(\ref{eq:rep_condition}), in the strictly
symmetric case ($n=1$), the GACE integrand is weight-independent, thus
leading to an ensemble $xc$ energy with weight $w$
that deviates from its ground state value by
$w[U-4t+\sqrt{U^2+16t^2}]/2$.
Therefore this deviation increases with the weight, as clearly
illustrated in Fig.~\ref{fig:eXC_energies}.
In the weakly correlated regime, the
deviation from the ground-state functional is essentially driven by the
exchange contribution, as expected. For $U=1$, the deviation induced by
the correlation energy becomes significant when approaching the
equiensemble case.
On the other hand, in stronger correlation regimes ($U=5$ and 10), the
weight-dependence of the ensemble correlation energy becomes crucial
even for relatively small ensemble weights. The bumps observed at $n=1$
are a pure ensemble correlation effect. In the light of Sec.~\ref{subsec:GACE_analysis}, we can conclude
that these bumps, which correspond to the largest deviation from the
ground-state {\it xc} functional, are induced by the plateaus in the
GACE integrand which are defined in the range $\vert 1-n\vert\leq \xi
$. Outside this range, the integrand is given by
Eq.~(\ref{eq:DD_fun_approx_outside_plateau}). Consequently, for given
ensemble weight $w$ and density $n$ such that $w\leq \vert 1-n\vert$, 
which leads to 
\begin{eqnarray}\label{eq:range_ens_GS_largeU}
w\leq n\leq 1-w \;\;\;{\rm or}\;\;\; 1+w\leq n\leq 2-w,
\end{eqnarray}
when
considering, in addition, the $v$-representability condition in
Eq.~(\ref{eq:cond_ens_dens}), 
the ensemble $xc$ energy (whose deviation from its ground-state value is obtained by integration from 0 to
$w$) can be approximated as follows,
\begin{eqnarray}\label{eq:ens_xc_ener_outside_plateau}
E^w_{xc}(n)&\approx&
E_{xc}(n)
+2t\Bigg(\sqrt{2}w+
\sqrt{(1-w)^2-(1-n)^2}
\nonumber\\
&&-\sqrt{1-(1-n)^2}\Bigg),
\end{eqnarray}
which approaches the ground-state $xc$ energy when $U/t\rightarrow+\infty$.
For finite but large $U/t$ values, we obtain at 
the border of the $v$-representable density domain ({\it i.e.} for $n=w$ or
$n=2-w$),
\begin{eqnarray}\label{eq:exc_ener_border}
E^w_{xc}(w)
&\approx& E_{xc}(w)
+\dfrac{2tw(3w-2)}{\sqrt{2}w+\sqrt{1-(1-w)^2}},
\end{eqnarray}
where the second term on the right-hand side is negative, 
and $E^w_{xc}(2-w)=E^w_{xc}(w)$ because of the
hole-particle symmetry. 
From these
derivations, 
we can match the behavior of the exact curves in Fig.~\ref{fig:eXC_energies}
for densities that fulfill Eq.~(\ref{eq:range_ens_GS_largeU}). Note finally
that, for such densities, the ensemble $xc$ potential can be
approximated as follows, according to Eq.~(\ref{eq:ehxc_pot_derivative})
and (\ref{eq:ens_xc_ener_outside_plateau}),
\begin{eqnarray}\label{eq:xcpot_outside_plateau}
&&\Delta v^w_{xc}(n)=-\dfrac{\partial E^w_{xc}(n)}{\partial n}
\nonumber\\
&&\approx
\Delta
v_{xc}(n)+
2t(n-1)\Bigg[\dfrac{1}{\sqrt{(1-w)^2-(1-n)^2}}
\nonumber\\
&&-\dfrac{1}{\sqrt{1-(1-n)^2}}\Bigg].
\end{eqnarray}
As expected and confirmed by the exact results of Fig.~\ref{fig:ens_xc_pot}, the ensemble
$xc$ potential becomes the ground-state one 
in the density domains of Eq.~(\ref{eq:range_ens_GS_largeU}) when $U/t\rightarrow+\infty$.\\
    
Let us now focus on the {\it complementary} range $w\geq \vert 1-n\vert$ or,
equivalently,
\begin{eqnarray}
1-w\leq n\leq 1+w.
\end{eqnarray}
In
this case, the ensemble $xc$ energy
is obtained by integrating over $[0,\vert 1-n\vert]$ and
$[\vert 1-n\vert,w]$ weight domains, thus leading to the following
approximate expression, according to
Eqs.~(\ref{eq:DD_plateau_exp}) and
(\ref{eq:DD_fun_approx_outside_plateau}), 
\begin{eqnarray}\label{eq:xcfun_ener_plateau}
&&E^w_{xc}(n)\approx
E_{xc}(n)+\dfrac{1}{2}\Big(U+\sqrt{U^2+16t^2}\Big)(w-\vert 1-n\vert)
\nonumber\\
&&+2t\Bigg(\sqrt{2}\vert1-n\vert+
\sqrt{(1-w)^2-(1-n)^2}
\nonumber\\
&&-\sqrt{1-(1-n)^2}\Bigg).
\end{eqnarray}
Turning to the ensemble $xc$ potential, it comes from
Eq. (\ref{eq:xcfun_ener_plateau}) that
\begin{eqnarray}\label{eq:xcpot_plateau}
&&\Delta v^w_{xc}(n)
\approx
\Delta
v_{xc}(n)
\nonumber\\+
&&
\Bigg[2t\sqrt{2}-\dfrac{1}{2}\Big(U+\sqrt{U^2+16t^2}\Big)\Bigg]\dfrac{\vert1-n\vert}{1-n}
\nonumber\\
&&+
 2t(n-1)\Bigg[\dfrac{1}{\sqrt{(1-w)^2-(1-n)^2}}
\nonumber\\
&&
-\dfrac{1}{\sqrt{1-(1-n)^2}}\Bigg].
\end{eqnarray}
Since, in the $U/t\rightarrow+\infty$ limit, the ground-state
$xc$ potential becomes discontinuous at $n=1$ and equal 
to~\cite{MP16_Senjean_2L-ILDA} 
\begin{eqnarray}\label{eq:xcpot_gs_stronglimit}
\Delta v_{xc}(n)\rightarrow 2U(n-1)+U\dfrac{\vert1-n\vert}{1-n},
\end{eqnarray}   
we conclude from Eq.~(\ref{eq:xcpot_plateau}) that, in the strongly correlated limit, the ensemble $xc$
potential becomes, in the range $1-w\leq n\leq 1+w$, 
\begin{eqnarray}\label{eq:exc_pot_strongcorr}
\Delta v^w_{xc}(n)\rightarrow 2U(n-1),
\end{eqnarray}
where, as readily seen, the
ground-state discontinuity at $n=1$ has been removed. This is in perfect agreement with the
$U=10$ panel of Fig.~\ref{fig:ens_xc_pot}. Note that, even though the
exact exchange potential varies also linearly with $n$, its slope is
weight-dependent (see Eq.~(\ref{eq:exact_ens_X_pot})) and equals the
expected $2U$ value only when $w=1/3$, as illustrated in
Fig.~\ref{fig:ens_xc_pot}. In other words, both exchange and correlation
contributions are important in the vicinity of $n$=1. Strong
correlation effects become even more visible at the borders of the bumps
in the $xc$ ensemble energy, namely $n=1\pm w$. Indeed, at these
particular densities, the ensemble $xc$ potential exhibits
discontinuities that are, according to
Eqs.~(\ref{eq:xcpot_outside_plateau}) and (\ref{eq:xcpot_plateau}),
equal to   
\begin{eqnarray}\label{eq:disc_ens_xc_pot_1pmw}
&&
\left.\Delta
v^w_{xc}(n)\right|_{n=(1\pm w)^+}
-\left.\Delta v^w_{xc}(n)\right|_{n=(1\pm w)^-}
\nonumber
\\
&&\approx
2t\sqrt{2}-\dfrac{1}{2}\Big(U+\sqrt{U^2+16t^2}\Big),
\end{eqnarray}
which becomes $-U$ when $U/t\rightarrow+\infty$. Let us stress that
Eq.~(\ref{eq:disc_ens_xc_pot_1pmw}) holds for $0<w\leq 1/2$. It relies
on the continuity of the ground-state $xc$ potential around $n=1\pm w$,
which explains why the ground-state case $w=0$ is excluded. Note finally
that, in the strongly correlated limit, the
ground-state discontinuity at $n=1$ equals, according to Eq.~(\ref{eq:xcpot_gs_stronglimit}),
\begin{eqnarray}
\left.\Delta
v_{xc}(n)\right|_{n=1^+}
-\left.\Delta v_{xc}(n)\right|_{n=1^-}=-2U,
\end{eqnarray}
which is twice the ensemble discontinuity at $n=1\pm w$, in  
agreement with the panel $U=10$ of Fig.~\ref{fig:ens_xc_pot}.

\section{Ground-state density-functional
approximations}\label{sec:GS_approx}

In practical eDFT calculations, it is common to use (weight-independent)
ground-state (GS) $xc$
functionals~\cite{senjean2015linear,PRA16_Alam_GIC}. Such an approximation
induces in principle both energy- and
density-driven errors. In this paper, we will only discuss the former,
which means that approximate ensemble energies are calculated with {\it
exact} ensemble densities.
The exact GS $xc$ functional will be used and the approximation
will be referred to as GS$xc$. 
The analysis of the density-driven errors ($\it i.e.$ the errors induced
by the self-consistent calculation of the ensemble density with the 
GS $xc$ density-functional potential) requires the use an
accurate parameterization for the GS correlation
functional~\cite{carrascal2015hubbard}. This is left for future work. For analysis
purposes, we also combined the exact (analytical) ensemble exchange
functional with the exact GS correlation functional, thus leading to the
GS$c$ approximation. In summary, for a given local potential $\Delta
v_{ext}$, 
the following exact
\begin{eqnarray}
E^w
&=&T^w_s(n^w)+(1-n^w)\Delta v_{ext}
+E_H(n^w)
\nonumber\\
&&+E^w_{xc}(n^w),
\end{eqnarray}
and approximate
\begin{eqnarray}
E_{{\rm GS}xc}^w
&=&E^w+E^{w=0}_{xc}(n^w)-E^{w}_{xc}(n^w),
\nonumber\\
E_{{\rm GS}c}^w
&=&E_{{\rm GS}xc}^w-E^{w=0}_{x}(n^w)+E^w_{x}(n^w)
\end{eqnarray}
ensemble energies have been computed, where $n^w$ is the {\it exact}
ensemble density. Note that, if Boltzmann weights were
used~\cite{PRA13_Pernal_srEDFT}, GS$xc$ would be similar to the
zero-temperature approximation (ZTA) of
Ref.~\cite{PRB16_Burke_thermal_DFT_dimer}. A significant difference,
though, is that ZTA is using a self-consistent density (thus inducing
density-driven errors) while, in GS$xc$, we use the exact ensemble
density. The comparison of GS$xc$, GS$c$ and ZTA is left for future
work.\\

\begin{figure}
\begin{center}
\includegraphics[width=1.0\textwidth]{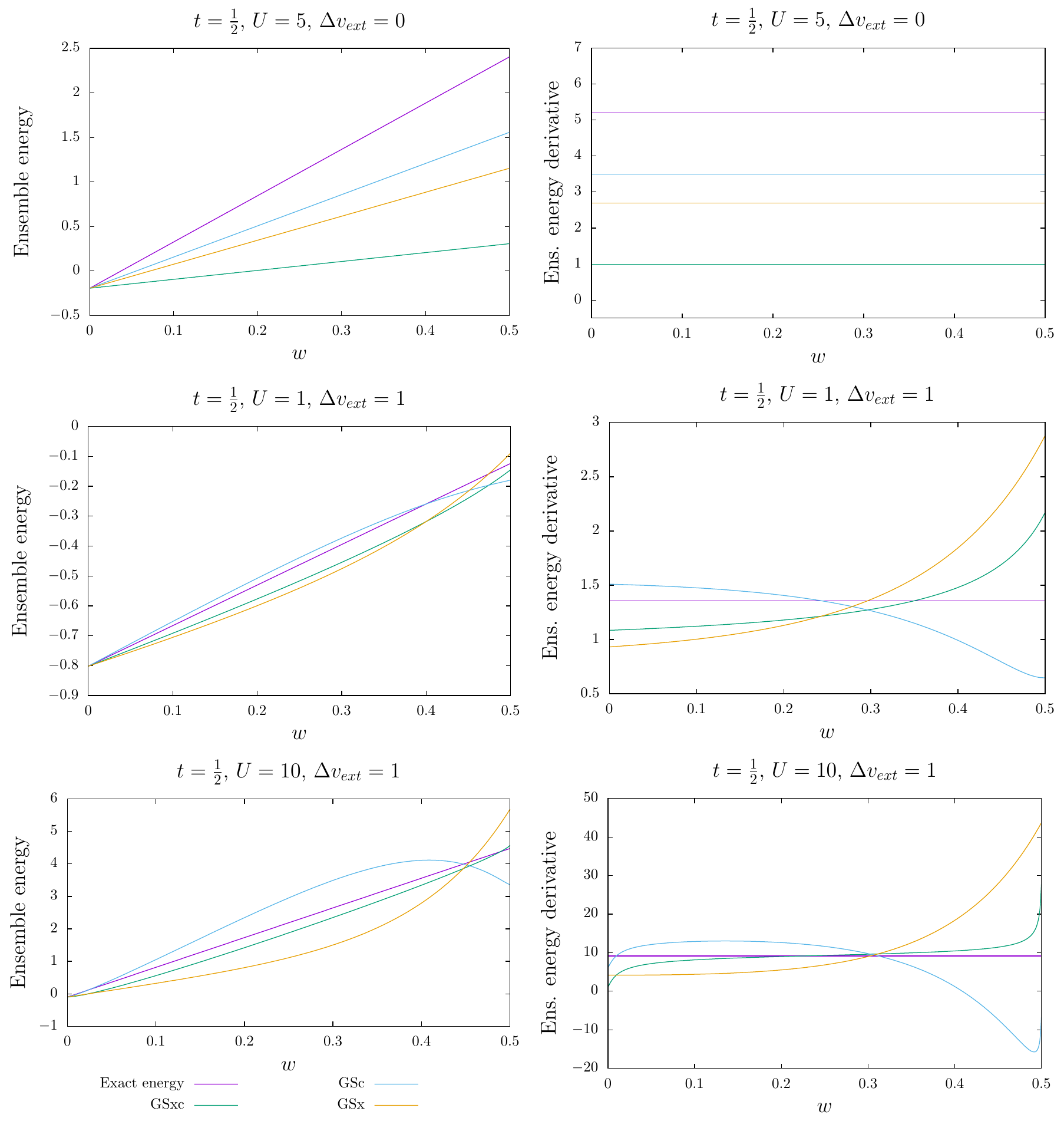}
\caption{Comparing exact with approximate ensemble energies (left panels)
and first-order derivatives (right panels) for various
$\Delta v_{ext}$ and $U$ values. See text for further details.}
\label{fig:approx_ens_energies}
\end{center}
\end{figure}

The approximate (weight-dependent) GS$xc$ and GS$c$ excitation energies are obtained by
differentiation with respect to $w$, thus leading to, according to
Eqs.~(\ref{eq:x_integrand_gace}), (\ref{eq:physpot_derivFw}),
(\ref{eq:deriv_Tsx_pot}) and (\ref{eq:exact_ens_X_pot}),
\begin{eqnarray}
&&\omega_{{\rm GS}xc}^w=
\left.\dfrac{\partial T^w_s(n)}{\partial
w}\right|_{n=n^w}
+\Bigg(\Delta v^{KS,w}(n^w)
\nonumber\\
&&-\Delta v^{KS,w=0}(n^w)
+\Delta v^{w=0}(n^w)
-\Delta
v_{ext}\Bigg)\dfrac{\partial n^w}{\partial w},
\end{eqnarray}
and 
\begin{eqnarray}
\omega_{{\rm GS}c}^w&=&\omega_{{\rm GS}xc}^w
+\Delta^w_{x}(n^w)
\nonumber\\
&&-\Big(\Delta v_x^w(n^w)-\Delta v_x^{w=0}(n^w)\Big)
\dfrac{\partial n^w}{\partial w}
,
\end{eqnarray}
where, according to Eqs.~(\ref{eq:Tsw}) and
(\ref{eq:true_phys_ensdensity0}),
\begin{eqnarray}
\left.\dfrac{\partial T^w_s(n)}{\partial
w}\right|_{n=n^w}&=&
\dfrac{2t(1-w)}{\sqrt{(1-w)^2-(1-n^w)^2}},
\\
\dfrac{\partial n^w}{\partial w}&=&n^1_{\Delta v_{ext}}-n^0_{\Delta
v_{ext}}.
\end{eqnarray}
Note finally that, when inserting the ensemble density of the KS system
$n^w=(1-w)n^{0,w}_{KS}+wn^{1,w}_{KS}$ into the Hartree
functional (see the first line of
Eq.~(\ref{eq:hartree_fun_def_hubbard_dimer})), we obtain the following
decomposition, 
\begin{eqnarray}
&&E_H(n^w)=(1-w)^2E_H(n^{0,w}_{KS})+w^2E_H(n^{1,w}_{KS})
\nonumber\\
&&+2Uw(1-w)\Big[1+(1-n^{0,w}_{KS})(1-n^{1,w}_{KS})\Big],
\end{eqnarray} 
where the last term on the right-hand side is an (unphysical)
interaction contribution to the ensemble energy 
that "couples" the ground and
first excited states. It is known as  
ghost-interaction error~\cite{PRL02_Gross_spurious_int_EDFT} and, since
$n^{1,w}_{KS}=1$ (see
Eq.~(\ref{eq:HFth_excited_dens})), it simply equals
$2Uw(1-w)$. This error is removed when employing 
the exact ensemble exchange functional, as readily seen   
in Eq.~(\ref{eq:Hubb_ens_EXX}). Therefore, GS$c$ is free from ghost
interaction errors whereas GS$xc$ is {\it not}. In the latter case, only half
of the error is actually removed, according to Eq.~(\ref{eq:x_fun_GS}). In order to visualize
the impact of the errors induced by the approximate calculation of the exchange energy
(which includes the ghost-interaction error), we combined the GS exchange functional with the exact ensemble correlation
one, thus leading to the GS$x$ approximate ensemble energy,
\begin{eqnarray}
E_{{\rm GS}x}^w
&=& E^w+E^{w=0}_{x}(n^w)-E^w_{x}(n^w)
\nonumber\\
&=&E^w+E_{{\rm GS}xc}^w-E_{{\rm GS}c}^w,
\end{eqnarray}
and the corresponding derivative,
\begin{eqnarray}
\omega_{{\rm GS}x}^w=\omega+\omega_{{\rm GS}xc}^w-\omega_{{\rm GS}c}^w.
\end{eqnarray} 
Results are shown in Fig.~\ref{fig:approx_ens_energies} for various
correlation regimes. In the symmetric case ($\Delta
v_{ext}=0$), $n^w=1$ so that both exact and approximate ensemble energies
are linear in $w$ and, as expected from Fig.~\ref{fig:eXC_energies}, GS$c$ performs
better than GS$xc$. In the asymmetric case ($\Delta v_{ext}=1$),
approximate ensemble energies become curved, as expected. For
$U=1$, GS$c$ remains more accurate than GS$xc$ (except for the
equiensemble). However, in the strongly correlated regime ($U=10$) and
for $w\geq 0.1$, the use of the ensemble exact exchange energy in
conjunction with the GS correlation functional induces large errors on the
ensemble energy. When approaching the equiensemble, the ensemble
energy becomes concave. The negative slope in the large-$w$ region leads
to negative approximate excitation energies, which is of course unphysical. On the other hand,
using both ground-state exchange and correlation functionals provides much
better results. This can be rationalized as follows. According to
Fig.~\ref{fig:densetat}, when $\Delta v_{ext}=1$ and $U=10$, the
equiensemble density equals 1.5, which corresponds to the border of the
bump in the ensemble $xc$ energy that was discussed previously. Using the $U=10$ panel of
Fig.~\ref{fig:eXC_energies}, we conclude that GS$c$ underestimates the
equiensemble correlation energy significantly while the exact ensemble
$xc$ energy is almost identical to the ground-state one. The former is
in fact slightly lower than the latter, as expected from
Eq.~(\ref{eq:exc_ener_border}) and confirmed by the $U=10$ panel of
Fig.~\ref{fig:approx_ens_energies}. Therefore, in this particular case,      
GS$xc$ is much more accurate than GS$c$. Interestingly, despite large 
errors in both exchange (which includes the ghost-interaction error)
and correlation energies for most weight values, relatively accurate
results are obtained through error cancellation. Note finally that, for $\Delta
v_{ext}=1$ and $U=10$, 
GS$xc$ and GS$c$ ensemble energy derivatives increase rapidly when
approaching the equiensemble case. This is due to the non-interacting
ensemble kinetic energy. Since the ground- and excited-state densities are
close to 1 and 2, respectively, $T_s^w(n^w)\approx -2t\sqrt{1-2w}$ and 
$\ddroit T_s^w(n^w)/\ddroit w\approx 2t/\sqrt{1-2w}$.           

\section{Conclusion}\label{sec:conclu}

eDFT is an exact time-independent
alternative to TD-DFT for the calculation of neutral excitation
energies. Even though the theory has been proposed almost thirty years ago, it is
still not standard due to the lack of reliable density-functional
approximations for ensembles. In this paper, exact two-state eDFT calculations
have been performed
for 
the nontrivial asymmetric two-electron Hubbard dimer. 
In this system, the density
is given by a single number which is the occupation $n$ ($0\leq n\leq 2$) of one of the two
sites. 
An exact analytical expression for the weight-dependent ensemble exchange energy has been
derived. Even though the ensemble correlation energy is not analytical,
it can be computed exactly, for example, by means of Legendre--Fenchel
transforms. 
Despite its simplicity, this model has shown many features
which can be observed in realistic electronic systems. In particular, the derivative discontinuity associated with
neutral excitations could be plotted and analyzed in various
correlation regimes. It appears that, in many situations, it is possible
to find an ensemble weight such that the KS gap equals exactly the
optical one.\\   
We have
also shown 
that, in order to connect the ensemble $xc$ functional with weight $w$
($0\leq w\leq 1/2$) to the
ground-state one ($w=0$), a generalized adiabatic connection for ensembles (GACE), where the
integration is performed over the ensemble weight rather than the
interaction strength, can be constructed exactly
for any ensemble-representable density. 
The GACE formalism was used for analyzing exact ensemble $xc$ energies
in the strongly correlated regime. In particular, we could show that
in the density domains $w\leq n\leq 1-w$ and $1+w\leq n\leq 2-w$, the
ensemble $xc$ energy is well approximated by the
ground-state one whereas, in the range $1-w \leq n \leq 1+w$, the ensemble and
ground-state $xc$ energies can differ substantially. The difference is
actually, in the strongly correlated limit, proportional to $Uw$ when $n=1$.
The existence of these three density domains is directly connected to 
the fact that, in
the strongly correlated regime, the well-known discontinuity at $n=1$ in
the ground-state $xc$ potential is removed when $w>0$ and it is replaced by two
discontinuities, at $n=1-w$ and
$n=1+w$, respectively.\\ 
Finally, ground-state density-functional approximations have
been tested and the associated functional-driven error has been
analyzed. Whereas the use of the exact (weight-dependent) ensemble exchange
functional in conjunction with the ground-state (weight-independent)
correlation functional provides better ensemble energies
(than when calculated with the ground-state $xc$ functional) in
the strictly symmetric or weakly correlated cases, the combination of
both ground-state exchange and correlation functionals provides much
better (sometimes almost exact) results away from the small-$w$ region
when the correlation is strong. Indeed, in the latter case, 
the ground-state density is close to 1 and the excitation corresponds to
a charge transfer, thus leading to an excited density close to 2 or 0. The
resulting ensemble density will therefore be close to $1+w$ or $1-w$. As
already mentioned, for $n=1\pm w$, the weight dependence of the
ensemble $xc$ functional becomes negligible as $U/t$ increases.  
This supports the idea that the use of ground-state functionals in
practical eDFT calculations
is not completely irrelevant. The analysis of density-driven errors is
currently in progress. 
One important conclusion of this work,
regarding its extension to {\it ab
initio} Hamiltonians, is that the
calculation of the GACE integrand plays a crucial role in the analysis
of exchange-correlation energies of ensembles and, consequently, in the 
construction of "true" approximate density functionals for ensembles.
The accurate computation of this integrand for small molecular systems
would be of high interest in this respect.      
We hope that the paper will stimulate new developments in eDFT. 

\section*{Acknowledgements}
The authors thank Bruno Senjean for fruitful discussions and the ANR (MCFUNEX project) for financial support.



\providecommand{\Aa}[0]{Aa}
%

\appendix*

\section{Energies and derivatives}

Individual ground- and first-excited-state singlet energies $E_i$ ($i=0,1$) are in
principle functions of $t$, $U$ and $\Delta v$, and they are solutions of 
\begin{eqnarray}\label{eq:3rdorder_energy}
-4 t^{2} U + (4 t^{2}- U^{2}+\Delta v ^{2})E_i+ 2 U E_i^{2} 
=E_i^{3}.
\end{eqnarray}
The exact 
ground-state energy can be expressed analytically as
follows~\cite{carrascal2015hubbard}, 
\begin{eqnarray}\label{eq:physical_energy_dimer}
E_0(U,\Delta v)=\dfrac{4t}{3}\left(u-w\,{\rm sin}\left(\theta +\dfrac{\pi}{6}\right)\right),
\end{eqnarray}
where 
\begin{eqnarray}
u=\dfrac{U}{2t},
\\ 
w=\sqrt{3(1+\nu^2)+u^2},
\\
\nu=\dfrac{\Delta v}{2t},
\end{eqnarray}
and
\begin{eqnarray}\label{eq:energy_phys_HUbb_end}
{\rm cos}(3\theta)=\left(9(\nu^2-1/2)-u^2\right)u/w^3.
\end{eqnarray}
The first-excited-state energy is then obtained by solving a
second-order polynomial equation for which analytical solutions can be
found~\cite{PRB16_Burke_thermal_DFT_dimer}.\\ 

Differentiating Eq.~(\ref{eq:3rdorder_energy}) with respect to $U$ gives
\begin{eqnarray} \frac{\partial E_i}{\partial U} = \frac{4 t^{2} + 2 U
E_i -
2 E_i^{2}}{4 t^{2} - U^{2} + 4 U E_i + \Delta v ^{2} - 3
E_i^{2}}.\label{hellfeyn}
\end{eqnarray}
Since, according to the Hellmann--Feynman theorem, 
\begin{eqnarray}
\bra{\Phi^{KS,w}_{i}(n)}\hat{U}\ket{\Phi^{KS,w}_{i}(n)}=U\left.\dfrac{\partial
E_i}{\partial U}\right|_{\Delta v^{KS,w}(n),U=0},
\end{eqnarray} 
combining Eqs.~(\ref{eq:Hubb_noint_E0}), (\ref{eq:Hubb_noint_E1}),
(\ref{eq:vksw_analytical})  with Eq.~(\ref{eq:Hubb_ens_EXX}) finally leads to the expression in
Eq.~(\ref{eq:Hubb_ens_EXX_ana}).\\
Similarly, we obtain the following expression for the derivative of individual energies with respect to the
local potential,   
\begin{eqnarray}\label{eq:deriv_ener_deltav}
\dfrac{\partial E_i}{\partial \Delta v}=\dfrac{2\Delta v
E_i}{3E_i^2-4UE_i+U^2-4t^2-\Delta v^2}.
\end{eqnarray}

\clearpage



\clearpage

\clearpage


\clearpage


\clearpage

\clearpage

\clearpage

\clearpage


\clearpage


\end{document}